\documentclass{aa}
\usepackage{graphics,natbib,amssymb}
\usepackage{rotating}
\newcommand{\solm}{M$_{\odot}$\ }
\newcommand{\solar}{L$_{\odot}$\ }

\newcommand{\rf}{\par\noindent\hangindent 15pt {}}
\begin{document}

\authorrunning{Moultaka et al.}
\titlerunning{Interstellar Matter and Circumstellar Matter in the
  Central Parsec of the Milky Way}
\title{Dust Embedded Sources at the Galactic Center}
\subtitle{2 to 4$\mu$m imaging and spectroscopy in the central parsec}
\author{J. Moultaka$^1$, A. Eckart$^1$, T. Viehmann$^1$, N. Mouawad$^1$,
C. Straubmeier$^1$, T. Ott$^2$, R. Sch\"odel$^1$}
\institute{1) I Physikalishes Institut,
           Z\"ulpicher Str. 77,
           50937 K\"oln, Germany \\
           2) Max Planck Institut f\"ur extraterrestrische Physik,
           Giessenbachstrasse, 85748 Garching, Germany\\
           \email{email: moultaka@ph1.uni-koeln.de} }

%\institute{

\date{Received  / Accepted }

\abstract{
\noindent

We present the first L-band spectroscopic observations for a dozen stellar sources in the central $0.5$~pc of the GC stellar cluster that are
bright in the 2-4 $\mu$m wavelength domain.  The L-band data were
taken with ISAAC at the VLT UT1 (Antu).  With the aid of additional
K-band spectroscopic data, we derive optical depth spectra of the
sources after fitting their continuum emission with a single reddened
blackbody continuum.  We also derive intrinsic source spectra by
correcting the line of sight extinction via the optical depth spectrum
of a late type star that is most likely not affected by local dust
emission or extinction at the Galactic Center. The good agreement
between the two approaches shows that the overall variation of the
line-of-sight extinction across the central $0.5$~pc is $\Delta
A_{\mathrm{K}}\leq0.5$~mag.  The extinction corrected spectra of the
hot He-stars are in good agreement with pure Rayleigh Jeans continuum
spectra.  The intrinsic spectra of all other sources are in agreement
with continuum emission and absorption features due to the dust
in which they are embedded.  We interprete both facts as evidence that
a significant amount of the absorption takes place within the central
parsec of the Galactic Center and is most likely associated with the
individual sources there. We find absorption features at $3.0\mu m$,
%$3.3\mu m$, 
 $3.4\mu m$, and $3.48\mu m$ wavelength. Correlations
between all %four 
three features show that they are very likely to arise in
the ISM of the central $0.5$~pc. 
%The $3.0\mu m$ ice feature appears to
%be related to material at higher temperatures (40~K) than found in
%earlier studies at larger distances from the center. 
Spacially highly
variable hydrogen emission lines seen towards the individual sources
give evidence of the complex density and temperature structure of the
mini-spiral. In the cases of the sources with featureless K-band
spectra like IRS~21 and IRS~1W, they are consistent with
massive hot stars embedded in the bow shock created by their motion
through the dust and gas of the mini-spiral.  It is likely that the
bow shock scenario may be applicable to most of the dust embedded
sources in the central stellar cluster. Spectroscopy of high
MIR-excess sources 0.5'' north of the IRS~13 complex is largely
consistent with them being YSOs. However, a bow-shock nature of these
sources cannot be excluded.  The L-band spectrum at the location of
SgrA*, is consistent with that of a hot O-type star, such as S2, which
was very close to Sgr~A* at the time of our observations.

}

\maketitle

%===================================================================

\section{Introduction}\label{sec:intro}

 Near-infrared diffraction limited imaging over the past 10 years
(Eckart \& Genzel, 1996; Genzel et al., 1997; Ghez et al., 1998, 2000;
Eckart et al., 2002; Sch\"odel et al. 2002, 2003; Ghez et al., 2003)
has resulted in convincing evidence for a 3-4$\times$10$^6$\solm black
hole at the center of the Milky Way.  This finding is supported by the
discovery of a variable X-ray and NIR source at the position of SgrA*
(Baganoff et al., 2001; Genzel et al., 2003a).  Most intrigingly,
near-infrared imaging and spectroscopic observations have provided
evidence for recent star formation in the central parsec of the Milky
Way, an environment previously thought hostile for star formation
because of the tidal field of the black hole, intense stellar winds,
and strong magnetic fields.

At a distance of 8~kpc (Eisenhauer et al. 2003), the Galactic Center is
surrounded by a circumnuclear ring of dense gas and dust showing
clumpy extinction (G\"usten et al. 1987). Inside this ring, there is a
central cavity of about 3~pc diameter, that contains mainly ionzed or
atomic gas. The visual extinction estimates towards prominent sources
within the central stellar cluster range between 20 and 50 magnitudes
with a median around 30 magnitudes (see Rieke, Rieke, \& Paul 1989,
Chan et al. 1997, Scoville et al. 2003).  In addition Scoville et
al. (2003) showed that the extinction is smoothly distributed across
the central 10 to 20 arcseconds with no indication of concentrations
of extinction on scales of about 1'' to 2''.

The $\sim$30 magnitudes of visual extinction along the line of sight
toward the Galactic Center (GC) is mostly due to the diffuse
interstellar medium (ISM)  (Lebofsky 1979) and in part due to
dense molecular gas (Gerakines et al. 1999; de Graauw et al. 1996;
Lutz et al. 1996).  The absorbing gas is cold (10~K) and the
abundances of important molecular species are similar to the solar
neighborhood (Moneti, Cernicharo, \& Pardo 2001a, Chiar et al. 2000).
In addition Blum et al. (1996) and Cl\'enet et al. (2001) concluded
that the colors of individual dusty sources within the central stellar
cluster contain a substantial contribution from intrinsic reddening.

The entire central parsec of our Galaxy is powered by
a cluster of young and massive stars (Blum et al. 1988, Krabbe et al. 1995,
Genzel et al. 1996, Eckart et al. 1999, Cl\'enet et al. 2001).
Within that cluster the 7 most luminous (L$>$10$^{5.75}$ \solar),
moderately hot (T$<$10$^{4.5}$~K) blue supergiants
contribute half of the ionizing luminosity of that region
(Najarro et al. 1997, Krabbe et al. 1995,  Blum et al. 1995).
Such massive and hot stars were also found in dense clusters
within the Galactic bulge, i.e. the Arches cluster (Cotera et al. 1992,
see also Figer et al. 2002 and references therein) and the
Quintuplet cluster (e.g. Figer et al. 1997).

In addition to the massive blue supergiants, a population of dusty
sources associated with bright dust emission can be found in the
Galactic Center stellar cluster.  After initial preceeding work by
Becklin \& Neugebauer (1968, 1969) first individual mid-infrared
sources in the central stellar cluster (among them IRS 1, 3 and
others) were reported by Rieke \& Low (1973) and Becklin \& Neugebauer
(1975).  Later, IRS 1 was resolved into multiple components by Storey
\& Allen (1983), Rieke et al. (1989), Simon et al. (1990), and Herbst
et al. (1993).  Further high resolution imaging by Tollestrup et
al. (1989) resolved IRS~6 and IRS~12 into multiple components. 

In this paper, we discuss MIR sources that are located well within the
central stellar cluster at projected distances from Sgr~A* of less
than 0.5~pc (Fig. \ref{slitpositions.eps}).  Several sources like
IRS~1, 3, and 21 are dominated by dust emission and are strong at a
wavelength of 10$\mu$m, whereas the supergiant IRS 7 is brightest at
2.2$\mu$m.  The nature of the dust enshrouded sources is still
unknown.  One among the best studied cases is IRS~21, which is
strongly polarized (17\% at 2$\mu$m ; Eckart et al. 1995, Ott et
al. 1999, Krabbe et al. 1995).  Initially, Gezari et al. (1985)
suggested that IRS~21 is an externally heated, high-density dust
clump.  Given the MIR excess and the featureless NIR spectra several
other classifications have been proposed, including an embedded
early-type star and a protostar (Blum et al. 1988, Krabbe et al. 1995,
Genzel et al. 1996, Cl\'enet et al. 2001).  Tanner et al. (2002)
suggest that IRS~21 is an optically thick dust shell surrounding a
mass-losing source, such as a dusty recently formed WC9 Wolf-Rayet
star.  Tanner et al. (2002, 2003) indicate that the extended dust
emission of most of the central sources is consistent with bow shocks
created by the motion of massive hot stars through the dust and gas of
the mini-spiral.

One way of investigating the nature of these bright NIR/MIR sources is 
imaging and spectroscopy in the 2 to 4$\mu$m wavelength range.
In addition to Hydrogen and Helium recombination lines
this wavelength domain is dominated by strong absorption features 
due to abundant molecules (NH$_3$, CH$_3$OH, H$_2$O, CO, CO$_2$ etc...), functional groups (like NH$_2$, CH$_2$), and ices.
Here H$_2$O ice enriched with molecular material is of special importance.
Liquid, crystalline, amorphous water ice as well as 
trapped water ice in SiO condensate (Wada et al. 1991)
give rise to a rich variety in shapes of a prominent feature 
with its deepest absorption at 2.94-3.00 $\mu$m
(e.g. Wada et al. 1991). The variety in shapes of the water ice feature is dependent not only on temperature, but also on annealing history and on the ice composition etc... (Hagen et al. 1983, Tielens \& Hagen 1982, Tielns et al. 1983, Kitta \& Kratschmer 1983, Hudgins et al. 1993, Maldoni et al. 1998) 

The emission of dust and the absorption features of ices are 
important diagnostic tools for the investigation of
the interstellar medium and circumstellar 
environments of individual sources.

Infrared sources towards the Galactic Center show a wealth of ice
absorption features (Butchart et al. 1986; Sandford et al. 1991)
indicative for a broad range of organic material mostly in the diffuse
interstellar medium.  Aliphatic hydrocarbons are characterized by
their CH$_2$ (methylene) and CH$_3$ (methyl) stretching modes around
3.4$\mu$m (Sandford et al. 1991; Pendleton et al.  1994).  Aromatic
hydrocarbons are detected via their CH and CC stretching modes at 3.28
and 6.2$\mu$m.  (Chiar et al. 2000, Pendleton et al. 1994).  An
absorption feature at 3.25$\mu$m has been found towards dense
molecular clouds. It is attributed to aromatic hydrocarbon molecules
at low temperatures (Sellgren et al. 1995; Brooke et al.  1999).
Differences in the exact central wavelength and profile width of the
absorption near 3.3$\mu$m are mostly attributed to differences in
temperature and/or carrier of the absorbing molecules in these
regions.

In this paper we present 3 to 4$\mu$m imaging and spectroscopy
data combined with near-infrared 2.2$\mu$m spectroscopy on the
strongest mid-infrared sources in the central stellar cluster. In addition to the previously published L-band observations of IRS~1W, IRS~3 and IRS~7 we provide the first L-band spectra of 9 other MIR sources: IRS~9, IRS~13, IRS~13N, IRS~21, IRS~29 and IRS~16~C, CC, NE and SW. These
data on sources located in the central $0.5$~pc of the GC enable us to study
the properties of the local interstellar medium and of circumstellar
matter in this region.

\section{Observations and data reduction}\label{sec:obs}

In order to investigate the nature of the dust embedded sources within 
the central 0.5~pc of the nuclear stellar cluster we used near- and 
mid-infrared imaging and spectroscopy. In the following we describe the
instrumentation and the data reduction that was employed.

\subsection{MIR Observations and data reduction}\label{sec:MIRobs}

The mid-infrared imaging and spectroscopy was obtained using the ISAAC
instrument at the ESO Very Large Telescope (VLT) unit telescope UT1
(Antu), at the Paranal observatory in Chile.  We have performed
spectroscopic and imaging observations of the Galactic Center during
May 23-30, 2002, as part of a monitoring program of SgrA* (Eckart et
al. 2003, Baganoff et al. 2003).  All the data reduction has been
performed using routines from the IRAF and MIDAS software packages.

For the imaging in the L-band, paired flat fielded images at different
chopper throws (18"$\pm$2") and chopping position angles (0$^o$ to
180$^o$) were subtracted from each other, resulting in frames
containing a positive and a (shifted) negative image.  The frames were
then shifted to a common reference point that coincides with a
positive image of a source.  Subsequently, frames belonging to the
same batch, i.e. taken sequentially with identical or different
chopper throws and/or chopping angles were combined by calculating a
median.  Since the images were moved to a common reference point this
procedure eliminates the negative "shadows" generated by the
subtractions.  This procedure also effectively removes cosmic rays and
bad pixel structures.  Such a batch typically consists of up to 40
images, and each resulting combined image covers an integration time
of approximately 35 minutes.  For the present investigation we used
the images with the best seeing and image quality with an angular
resolution of about 0.4''.  The absolute L-band flux calibration (see
magnitudes listed in Table \ref{tabfit}) was performed using the
fluxes of several bright well isolated objects also measured by
Cl\'enet et al. (2001). 

\begin{figure}
  \resizebox{8.5cm}{!}{\rotatebox{0}{\includegraphics{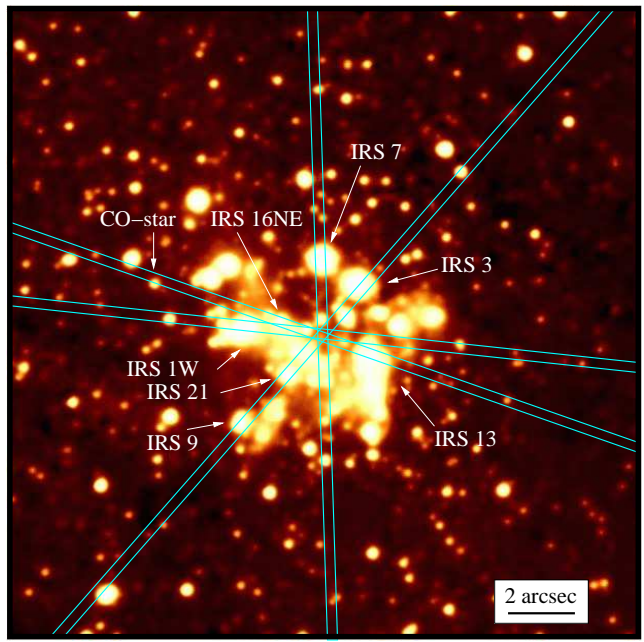}}}
  \caption[]{L-band image as obtained with ISAAC at the VLT UT1. 
   The four slit positions chosen for the spectroscopic observations 
   are also shown. 
   They all run through the position of SgrA*. 
   The position of the CO-star used for the calibration along the 
   line of sight is shown as well.}
\label{slitpositions.eps}
\end{figure}

The spectroscopic observations were performed with the long-wavelength
(LWS3) and low resolution (LW) mode using the SL filter covering the wavelength range of 2.7$\mu$m - 4.2$\mu$m and
4.4$\mu$m - 5.1$\mu$m, respectively.  The use of a 0.6'' slit width
implied a spectral resolution ($R=\lambda/\Delta \lambda$) of R=600 in
that wavelength domain.  The seeing at this time was ranging between
0.4 and 0.9 arcseconds.  In order to compensate for the thermal
background separate chopped observations were carried out using
chopper throws of $\sim$18 arcseconds and random nodding within 2
arcseconds along the slit.  We adopted 4 different slit positions all
running through the $Sgr A^*$ location (see Fig. \ref{slitpositions.eps}).  The resulting images were divided by
flat-fields, corrected for cosmic rays, for sky lines and dispersion
related distortion.  The wavelength calibration has been performed
using a Xenon-Argon lamp. 

Two chopped frames (with shifted image positions) were then subtracted
one from each other to provide a single frame containing two negative
trace images and a positive one with twice the intensity than the
negative images.  After extraction of the individual source spectra,
they were corrected for wavelength dependent sensitivity, atmospheric
transmission, and telluric lines using two standard stars HD 194636
(B4V) and HD 148703 (B2III-IV).

The spectra were normalized to the given K- and L-band magnitudes
listed in Table \ref{tabfit}.

\subsection{NIR Imaging spectroscopy and data reduction}\label{sec:NIRobs}

The near-infrared data in the K-band (2.2$\mu$m) was obtained via 
integral field spectroscopy using the imaging spectrograph
3D (Weitzel et al., 1996) combined with the tip/tilt corrector ROGUE
(Thatte et al., 1995) at the ESO/MPG 2.2m located at La Silla.
This instrument allows observations of a continuous 2-dimensional
field ($16\times 16$ pixels) while providing spectral information
for each spatial image element. 
These seeing-limited observations were done
with a pixel scale of $0.3``$ resulting in a field
of view of $4.8``\times 4.8``$ for individual pointings.
Using a spectral resolution of $R=2000$ 
each half of the K-band had to be covered separately.

The observations were centered on the IRS~16 cluster.
In March 1996, the central parsec was observed with a spectral coverage
from $1.97$ to 2.21$\mu$m (lower half of the K-band). 
In total 17 different but overlapping regions were observed.
The total sky area covered amounts to $16\arcsec $ in East-West
and $10\arcsec $ in North-South direction.
The upper half of the K-Band from $2.18$ to 2.45$\mu$m was
observed during a second observing run in April 1996. 
Here a total of 52 overlapping regions was observed.
This resulted in an area of $25\arcsec $ in North-South direction 
and $20\arcsec $ in East-West. 
For further details of the observations see 
(Genzel et al. 1997, Ott et al. 2003).

In order to calibrate the wavelength scale,  spectral lamp data 
(argon lamps in this case) were taken at the beginning or the
end of each observing night.
As a further calibration step, calibration sources with a known spectrum
were observed at a similar airmass as the Galactic Center. These standard
stars were divided by a spectrum of the same stellar type (Kleinmann
\& Hall 1986) in order to remove stellar features resulting in an 
atmospheric transmission spectrum. 
The source data were then divided by this spectrum.

%===================================================================
\section{Extinction correction}

The patchy extinction towards the GC stellar cluster (see e.g.\
Scoville et al. 2003) demand a careful calibration of spectral
data. Therefore, the correction of the extinction along the line of sight
has been carried out using the method described in Sect. \ref{sec:Fitting} and tested using another, independent method described in Sect. \ref{sec:contr}. 
%In our work, we adopted a double approach to this task. On the
%one hand, we used the spectral continuum at selected wavelength ranges
%in order to fit reddened blackbody emission to the spectra. On the
%other hand, 
%we used a CO star that can be considered free of local
%absorption for the derivation of the wavelength dependent
%line-of-sight extinction. 
A cross-check shows that both of our
approaches agree very well (see the results in Sect. \ref{sec:rescontr} and the conclusion (Sect. \ref{sec:conclusion}) at the end of this section) and that the extraction of the optical depth spectra described in Sect. \ref{sec:opticaldepth} can be done safely.

\subsection{Fitting Blackbody Emission to the 2 to 4$\mu$m Spectral Continua}\label{sec:Fitting}

\subsubsection{Method}

In order to estimate the extinction towards the individual sources and
to determine the approximate continuum shape of the observed spectra,
we have performed simultaneous fits of our K- and L-band data with
single, reddened blackbody spectra.  In this process we have
considered the continuum emission, fitted locally around $2.2\mu m$,
$3.75\mu m$, and $4.17\mu m$ wavelength, as representing the
intrinsic, reddened continuum of the sources.  In order to find the
best fit, the temperature of the blackbody was allowed to vary within
limits with a step size of $100~K$.  For the Helium stars these limits
were 20,000 and 26,000~K (Najarro et al. 1997).  For the remaining
objects the limits were 200~K and 4,000~K.  We allowed the K-band
absorption $A_K$ to vary over a range between 2.7 to 4.5 magnitudes with
a step sizes of $0.05\,mag$ assuming the extinction law $A_\lambda
\sim \lambda^{-1.8}$ stated by Martin \& Whittet (1990). 
We extended the fitting range towards higher $A_K$ values in order to
allow a search for higher extinction towards the dusty sources.
In general the fitting is not very sensitive with respect to the 
determination of $A_K$ due to the strong H$_2$O absorption 
feature that dominates almost the entire L-band.
%The resulting fits to the individual spectra are shown in Figs.
%\ref{Fitspectra1.eps} to \ref{Fitspectra3.eps}.

There are several potential problems that one has to consider when
using this procedure (see also discussion in Chiar et al. 2002):

We have to take into account that our K- and L-band data have not
been taken at the same epochs and in some cases their relative
calibration may be affected by variability.  Some sources in the
central parsec of the Milky Way are variable on time scales of months
to years (Ott et al. 1999, Blum et al. 1996, and Tamura et al. 1994,
1996).  In order to achieve a successful combined fit over both bands
we therefore used different overall flux calibrations of the spectra.
For the K-band fluxes, we have considered data from 5 different
references: Becklin et al. (1978), Tollestrup et al. (1989), Blum et
al. (1996), Ott et al. (1999), and Cl\'enet et al. (2001).  The
observations of Cl\'enet et al. (2001) are the most recent of the four
references and have been obtained with Adaptive Optics.  For this
reason, we have privileged the use of the values given in Cl\'enet et
al. (2001) for the fitting procedure.  However, for some sources, a
better fit was obtained when using a different flux calibration.  This
was the case for IRS 7, IRS 9, the stars in the IRS 16 complex as well
as IRS 13.  In the case of IRS 7 and IRS 9, these objects are variable
as shown in Ott et al. (1999), Blum et al. (1996), and Tamura et
al. (1994, 1996).  The relative flux density calibration from Becklin
et al. (1978) results in the best fits for our data since their
observations in the K- and L- bands were performed in the same year.
The IRS~13 complex consists of at least three bright members, E1, E2,
and E3 (Paumard et al. 2001, Maillard et al. 2003).  They contain the
hottest and most luminous star in the entire region (Najarro et
al. 1997). The IRS~13 complex therefore is very likely to be variable
in flux density.  In this case the fitting results were best using the
K-band fluxes by Blum et al. (1996). \\
The blackbody temperatures and K-band extinction resulting from the best fit of the spectra are listed in Table~\ref{tabfit}.

\begin{figure*}
  \resizebox{15cm}{!}{{\includegraphics{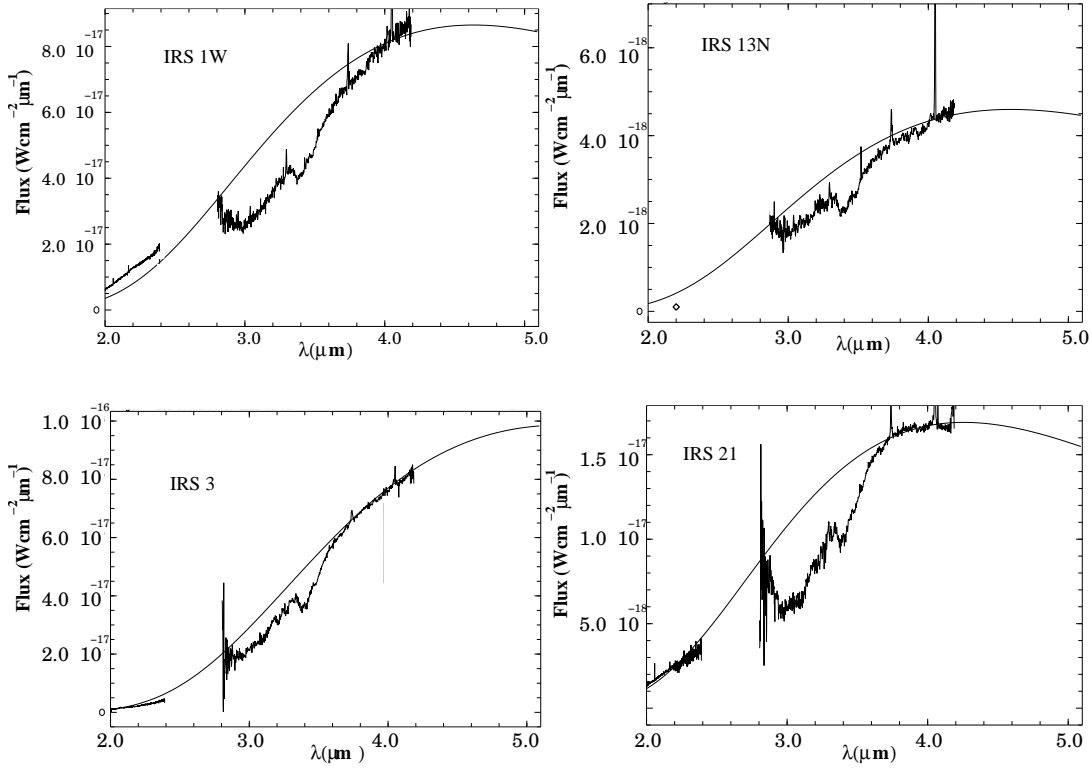}}}
  \caption[]{Best fits of the K- and L- band spectra with single 
reddened blackbody continua. When no K-band spectrum was available, 
the flux density at $2.2\mu m$ represented by a diamond was 
considered in the fitting procedure.}
\label{Fitspectra1.eps}
\end{figure*}

\begin{figure*}
  \resizebox{15cm}{!}{{\includegraphics{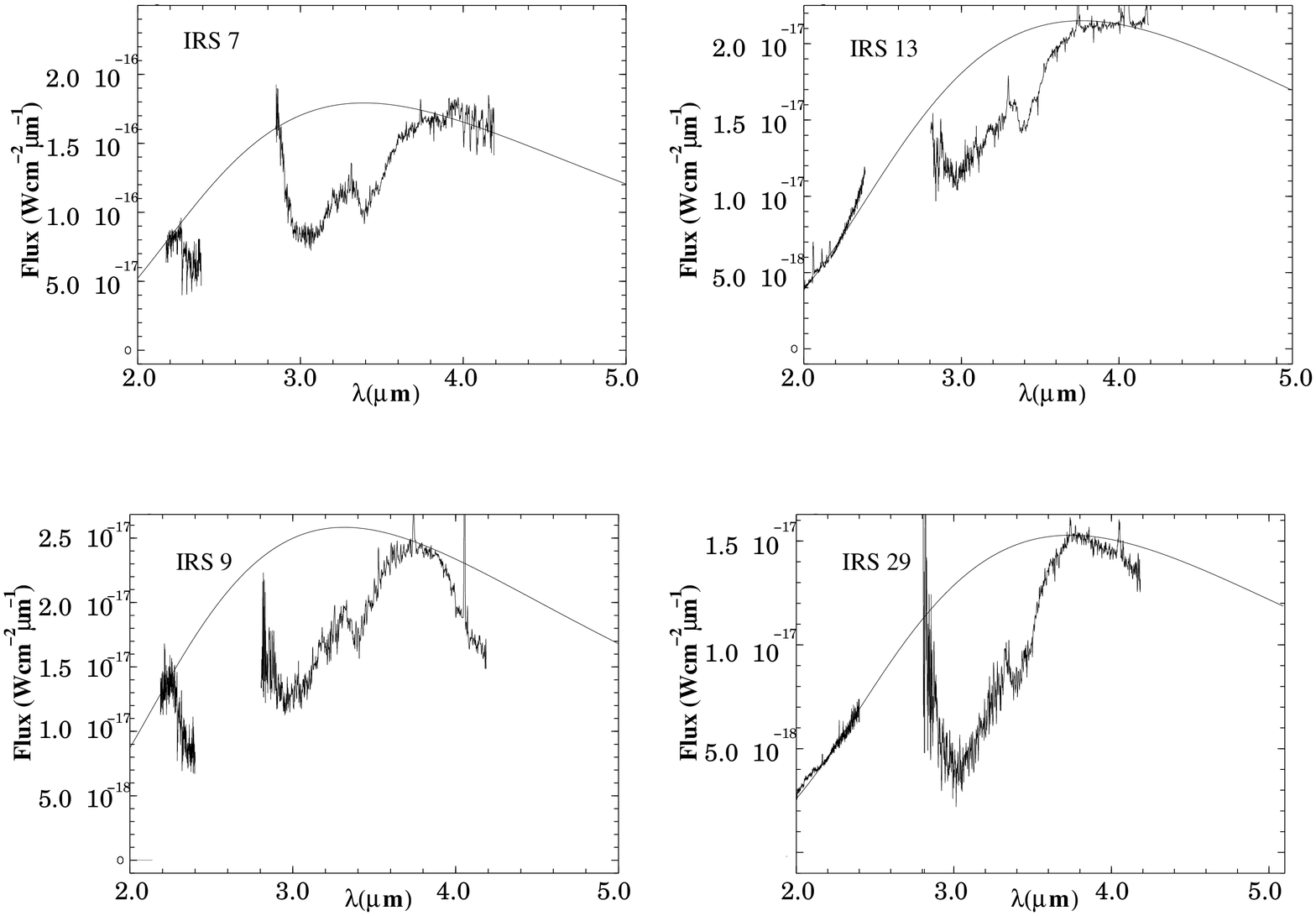}}}
  \caption[]{Fitting spectra with single reddened blackbody continua (cont.).}
\label{Fitspectra2.eps}
\end{figure*}

\begin{figure*}
  \resizebox{15cm}{!}{{\includegraphics{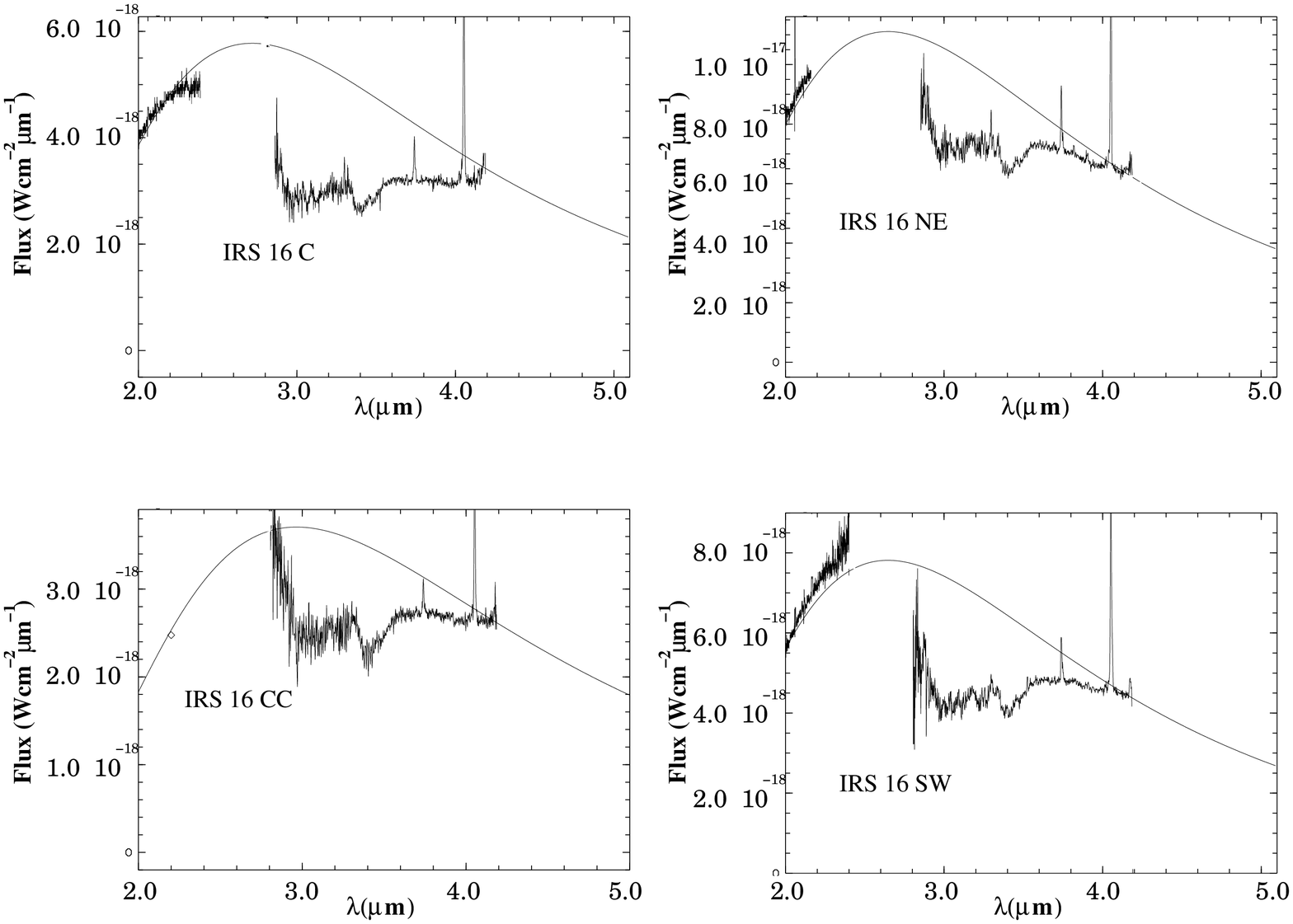}}}
  \caption[]{Fitting spectra with single reddened blackbody continua (cont.).}
\label{Fitspectra3.eps}
\end{figure*}

Moreover, most of the spectra are heavily affected by broad absorption 
features and it is difficult to determine a clear measure of the underlying
continuum emission. Also, the assumption of a single temperature
blackbody continuum can only be taken as a first approximation. On the
other hand, reddened multi-temperature models quickly result in a
larger number of not well determined parameters (e.g. temperature and
relative flux density contribution for each component).\\
The resulting best fits of the individual spectra with a single blackbody are shown in Figs~\ref{Fitspectra1.eps} to \ref{Fitspectra3.eps}.

\begin{table*}[htbp]
\small
\begin{center}
\begin{tabular}{|r|c|c|c|}
\hline
Source & Blackbody  & $A_k$  & L-band \\
 & Temperature &   & magnitude \\
\hline
IRS 1W& 900 & 3.30 & 5.8 \\
IRS 3  & 800 & 4.20 & 5.3 \\
IRS 7  & 2100& 3.55  &5.0 \\
IRS 9  & 1900& 3.10  &7.4 \\
IRS 21 & 1200& 3.95 &6.8 \\
IRS 29 & 1600& 3.70  &7.4 \\
IRS 16C& 22600 & 3.45 &8.6 \\
IRS 16CC& 20200& 4.00 &9.0 \\
IRS 16NE& 22400& 3.25 &7.7 \\
IRS 16SW& 22500 & 3.25  &8.4 \\
IRS 13& 1300 & 3.00 & 6.7\\
IRS 13N& 1000& 3.90 & 9.76\\
\hline
\end{tabular}
\end{center}
\caption{Results from fitting a single reddened blackbody to the K- and L-band spectra.
In addition to the blackbody temperature and the K-band extinction obtained from the fitting procedure, we list the L-band 
magnitudes ($\pm$0.2 mag uncertainty) we derived from our images.
}
\label{tabfit}
\end{table*}

\subsubsection{Results}\label{fitresults}

We list the resulting parameters of the fits in Table~\ref{tabfit} and
the spectra with the corresponding blackbody curves in
Figs.~\ref{Fitspectra1.eps}, \ref{Fitspectra2.eps}, and
\ref{Fitspectra3.eps}. The temperatures and the K-band extinctions
derived by the fitting procedure agree well with those found by other
authors: The He stars of the IRS 16 complex are best fitted with
temperatures that are similar to those found by Najarro et al. (1997).
IRS~7 and IRS~9 have typical temperatures of late-type supergiants or
giant stars (Chiar et al. 2002, Ott et al 1999) and the remaining
dusty sources are well fitted by blackbody continua of typical
temperatures for hot dust (Tanner et al. 2002, Gezari et al. 1996,
Genzel et al. 1997, Blum et al. 1996 and others).
% 
%The single blackbody approach thus appears to be responsible that in 
%some cases (e.g. IRS~7, IRS~29 in Figs. \ref{Fitspectra1.eps} and
%\ref{Fitspectra2.eps}) the top of the blue wing in the 3$\mu$m ice
%absorption, the individual K-band (in the case of IRS~1W shown in Fig.
%\ref{Fitspectra1.eps}) or the L-band (in all IRS~16 spectra) continuum
%levels and shapes are not fitted very well. However, 

%Moreover, the {\it overall} shapes we obtain by fitting single 
%blackbodies are reasonable. They provide an improvement over the study made by Sandford et al. (1991) where the authors used a linear continuum between $2.8\mu m$ and $3.7\mu m$ as baseline to derive the optical depth spectra of the 2 sources in common IRS~3 and IRS~7. Their choice was imposed by the reduced wavelength range of their data. Here, our data span a larger domain from $2.0\mu m$ up to $4.2\mu m$. Unlike the study of Sandford et al. (1991), we used the complete available information on the extinction from $2\mu m$ to $4.2\mu m$ toward the Galactic Center. \\
The overall shapes we obtain by fitting single blackbodies are reasonable and compare favorably to those of Chiar et al. 
(2002) (especially in the case of the three sources in common IRS~1W, IRS~3 and IRS~7). In addition, they are supported by an independent procedure to calibrate the line of
sight absorption described in the following section (Sect. \ref{sec:contr}).
 Also, the inclusion of the K-band spectra which was missing in the work by Chiar et al. (2002) %represents an improvement over the
%work by Chiar et al. (2002). It 
allows to better judge the quality of the
fitting procedure via comparison to the K-band continuum fluxes and
spectral shapes, especially for the three sources in common (IRS~1W, IRS~3 
and IRS~7).

Despite of the improvements in the spectral fitting, there are some mismatches we want to comment on:\\
The fit of the IRS~1W spectrum shown in Fig.~\ref{Fitspectra1.eps} does not 
match the K-band spectrum of this source and shows a continuum level mismatch in the $3.8\mu m$ to $4.0\mu m$ region
 that has no physical significance. This is certainly due to a non-consistency 
between the flux calibrations in the K- and L- spectral bands.
 For this reason, we decided to set an upper and a lower limit to the fit by
 performing a fit matching perfectly the K-band spectrum without caring about the adjustement at the red part of the spectrum on the one hand 
and another fit compensating the non physical mismatch at $3.8\mu$m.

An absorption at $\sim 4.2\mu$m appears in the spectra of IRS~9 and IRS~29. 
This suppression in flux is hardly an artefact of the data reduction, since we used the same procedures as for the other sources. We have no scientific explanation for that suppression but considered it as real. Also, both in shape and center 
wavelength, it does not seem to be consistent with the closest (at $4.27\mu m$) strong 
absorption line due to the stretching mode resonance of solid $CO_2$ also 
observed towards the dust shells of some Young Stellar Objects (de Graauw et
al. 1996). The presence of this suppression does not affect results presented and conclusions drawn in this paper.
  
Concerning the IRS 16 objects, all fits were not very satisfying
towards the red part of the L-band spectrum and match perfectly well the K-band spectra. It is obvious that this is not due to the fitting procedure itself but to the shape of the L-band spectra of these sources which are very flat (except for the IRS~16~NE case where the shape matches better the overall shape of a Rayleigh-Jeans spectrum). These spectra have been reduced in the same way as all the other spectra with the same flux calibrator stars and thus the shape of the spectra is real.\\ 
The origin of this behavior is not clear. It may be due to the fact
that these sources are the least contaminated by local dust emission
features and are likely most susceptible to variations in the
wavelength dependent line of sight extinction or of properties of the
associated material. If for instance - compared to the intrinsic,
local absorptions of the dusty sources - the line of sight absorption
is dominated by amorphous H$_2$O ice (Wada et al. 1991), the
corresponding line center and strength of the red wing would be
shifted towards the red. As a result the slope of the reddened
continuum simultaneously fitting the K- and L-band continuum would be
systematically too large.

\begin{figure*}
  \resizebox{16cm}{!}{\rotatebox{90}{\includegraphics{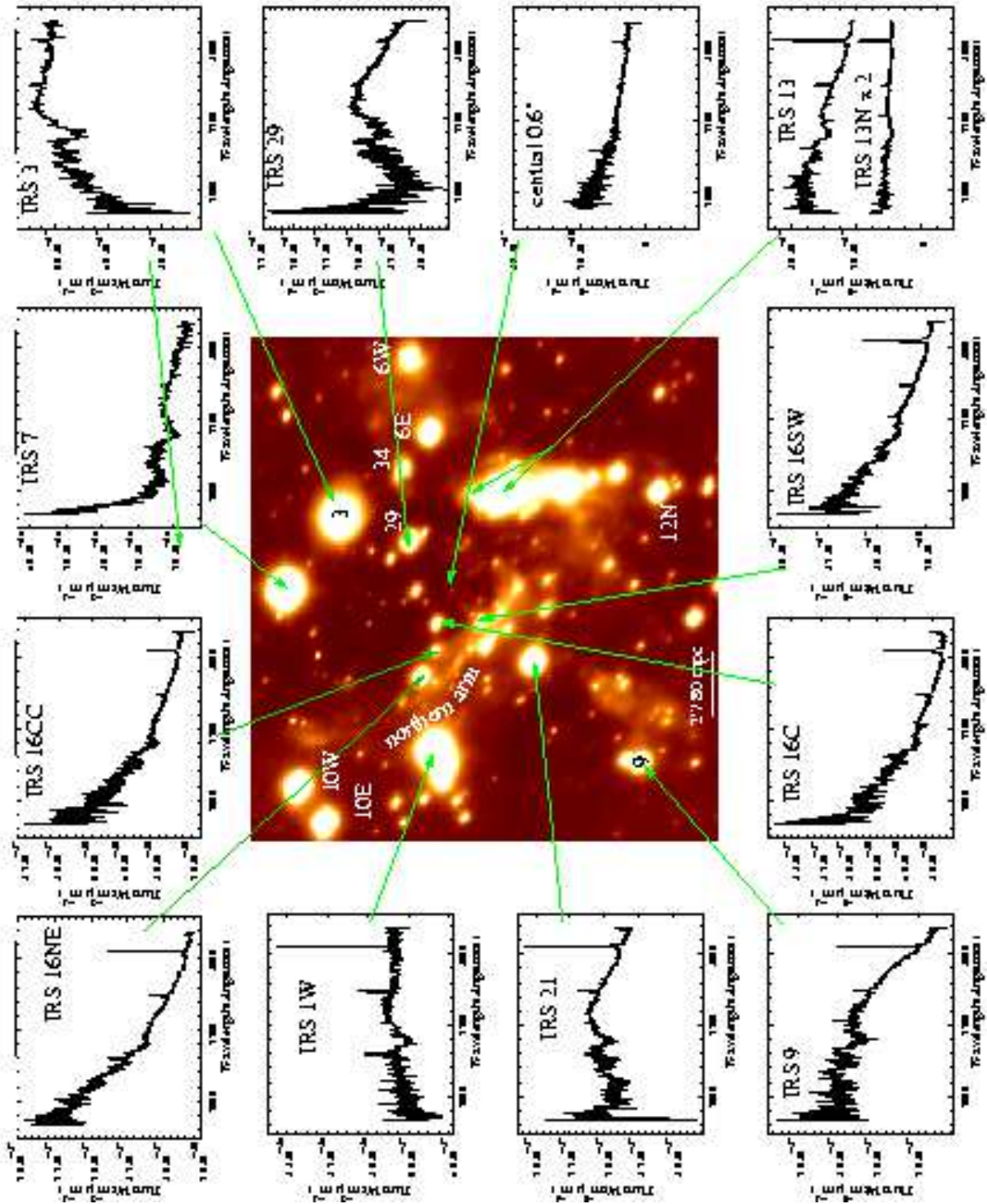}}}
  \caption[]{Our ISAAC L-band map combined with the spectra of the 
   central sources corrected for the line of sight extinction.}
\label{lbandmos2.eps}
\end{figure*}

\subsection{Determining the Wavelength-Dependent Line of Sight Extinction} \label{sec:contr}

\subsubsection{Method}

The calibration procedure described in this section is supported by
the finding of Scoville et al. (2003) that the extinction shows a
smooth distribution across the central 10 to 20 arcseconds with no
indication of concentrations of extinction on scales of about 1'' to
2''. Furthermore Blum et al. (1996) and Cl\'enet et al. (2001)
concluded that the colors of the sources within the central stellar
cluster are due to both, intrinsic and foreground reddening.\\

\begin{figure}
  \resizebox{7cm}{!}{\rotatebox{0}{\includegraphics{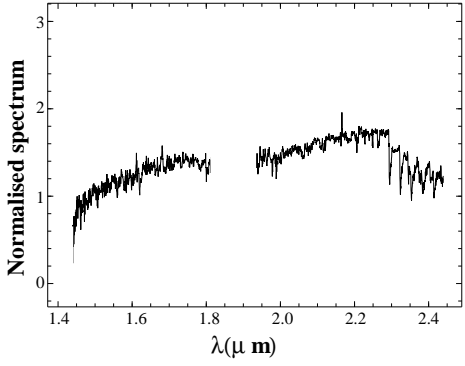}}}
  \caption[]{The H- and K-band spectrum of the late-type CO-star of which L-band spectrum is used for the line of sight absorption correction.}
\label{COstarKband}
\end{figure}

\begin{figure}
  \resizebox{7cm}{!}{\rotatebox{0}{\includegraphics{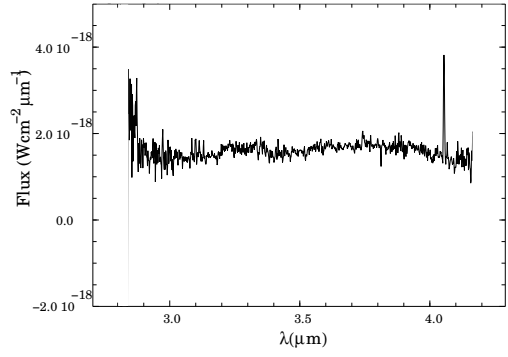}}}
  \caption[]{The L-band spectrum of the late-type CO-star used for the line of sight absorption correction.}
\label{COstar.eps}
\end{figure}

One of the slit settings we adopted for our observations runs through
a late-type star shown in Fig. \ref{slitpositions.eps}.  This star
shows clear 2.3$\mu$m CO absorption band-heads in its K-band spectrum (see Fig. \ref{COstarKband}), therefore, we will call it hereafter ``CO-star'' (e.g. Eckart et al. 1995).
 It is located at a projected distance of 12.6`` ($\sim 0.5$ pc)
from the center, is well off the mini-spiral emission and does not
show excess emission at wavelengths of 3$\mu$m or longer (see the L-band spectrum in Fig. \ref{COstar.eps}).\\

Therefore we can safely assume that  this star is largely free of {\em
local} reddening and that its spectrum is mostly affected by the line
of sight extinction along the 8~kpc towards the Galactic Center.\\

We assume that the L-band spectrum of the CO-star can be represented by a
3600~K (M0-M3 spectral type) blackbody spectrum.  If one matches the $\sim$4.2$\mu$m flux
densities of the measured L-band spectrum and the theoretical
continuum blackbody spectrum then the ratio between both spectra
provides a measure of the relative wavelength dependent ice feature
dominated L-band absorption due to the interstellar medium along the
line of sight.  The corresponding optical depth spectrum is shown in
Fig. \ref{Taulineofsight4.eps}.  If we correct this spectrum with
the continuum optical depth value at $4.2\mu m$ of $\tau$=1.09
(derived via interpolation between the expected L- and M- band
extinctions given by Rieke \& Lebofsky 1985) the mean optical depth
would be of the order of $1.64$.  This result is consistent with the
value of $1.55$ obtained by Rieke \& Lebofsky (1985) for the L-band
optical depth.  We have divided the spectra of all remaining objects
by this extinction spectrum (of which the optical depth spectrum is
shown in Fig. \ref{Taulineofsight4.eps}).  The resulting spectra
then represent probably more closely the source spectra as seen at the location of the
Galactic Center - corrected for wavelength dependent absorption along
the line of sight towards the Galactic Center.  

\subsubsection{Results}\label{sec:rescontr}

\begin{figure*}
  \resizebox{15cm}{!}{{\includegraphics{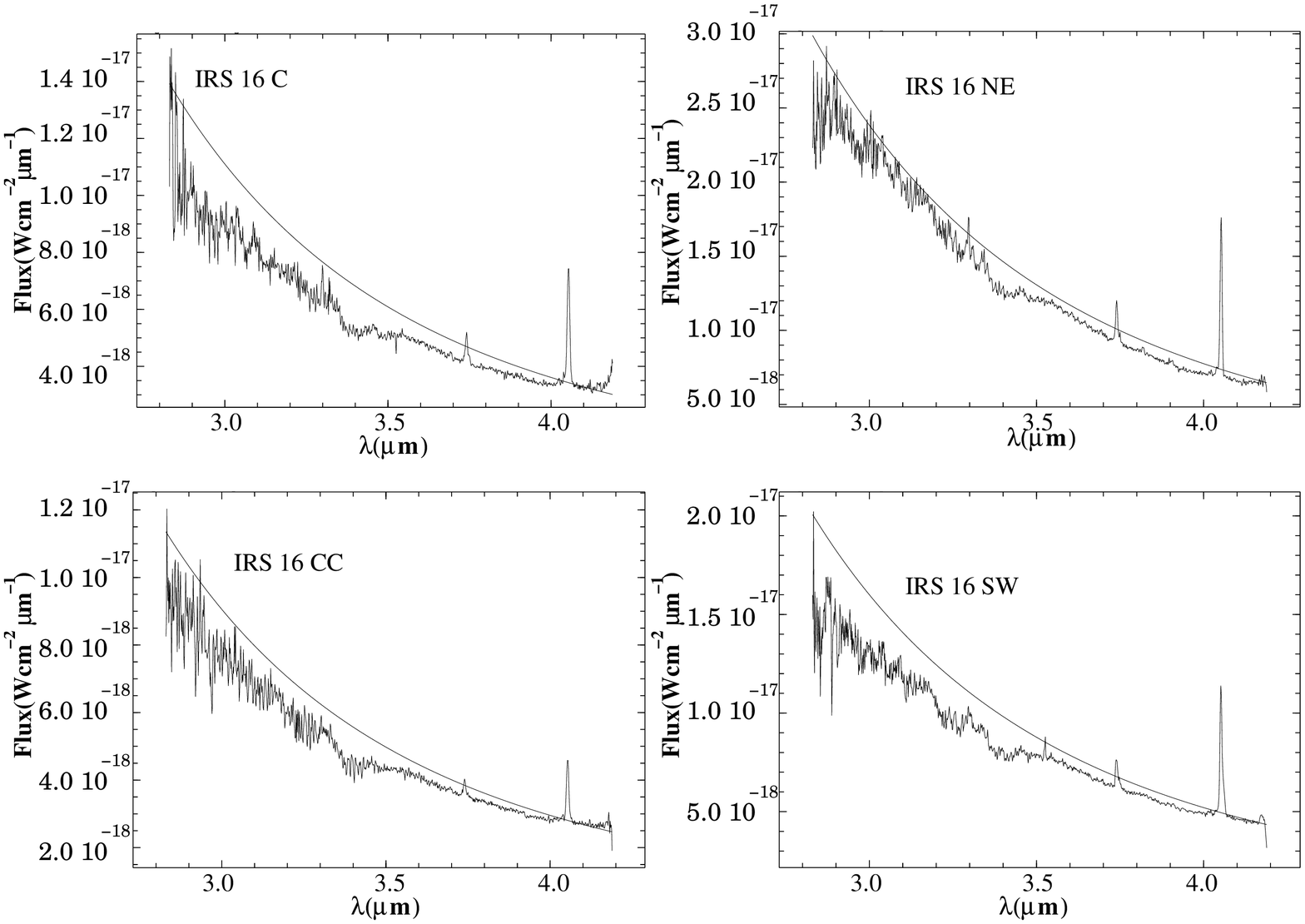}}}
  \caption[]{L-band spectra corrected for line of sight absorption (cont.).}
\label{Extcorr3.eps}
\end{figure*}

We show the spectra after correction for wavelength-dependent
absorption together with our L-band image in
Fig.~\ref{lbandmos2.eps}.\\
In Figs.~\ref{Extcorr1.eps} to \ref{Extcorr3.eps}, the spectra
corrected for the measured wavelength-dependent line of sight
extinction as determined from the CO-star are shown in comparison to
the continuum spectra that were derived by fitting the {\it
non}-corrected spectra with a single reddened blackbody
(Sect.~\ref{sec:Fitting}).\\
We find that they can be grouped into
three classes represented by the three figures.

\begin{itemize}
\item I. As shown in Fig.~\ref{Extcorr3.eps}, the spectral shapes of
the hot stars found in the IRS16 complex are fitted well by Rayleigh
Jeans continuum spectra i.e.  high temperature blackbody spectra (listed in Table \ref{tabfit}).\\
The slight mismatch between the blackbody
spectra and the calibrated observed spectra in the red part of the
L-band wavelength range corresponds to the one between the uncorrected spectra
and the reddened blackbody continua shown in
Fig.~\ref{Fitspectra3.eps}. Its origin is, thus, probably the same as mentioned in the previous section.

\item II. The corrected spectra of IRS~1W, IRS~3, IRS~13N and IRS~21
(Fig.~\ref{Extcorr1.eps}) are flat in the red part with indications
of continuum absorption at wavelengths short-ward of $3.2\mu m$. The
spectra can be roughly explained by a T$\le$1500~K blackbody
continuum, with deviations at wavelengths $\leq3.5\mathrm{\mu m}$ due
to the red wing of the $3.0\mathrm{\mu m}$ ice absorption feature.

\item III. Finally, in addition to strong H$_2$O ice absorption the spectra
of IRS~7, IRS~9, IRS~13, and IRS~29 (Fig. \ref{Extcorr2.eps}), show
long-ward of $3.2\mu m$ the characteristic shape of a T$\sim1500~K$ to
$2000~K$ continuum. The absorption at $4.2\mu m$ corresponds to that distinguished in the observed spectrum of Fig.~\ref{Fitspectra2.eps}. 
\end{itemize}

\subsection{Conclusion on the extinction correction}\label{sec:conclusion}

\begin{figure}
  \resizebox{7cm}{!}{\rotatebox{0}{\includegraphics{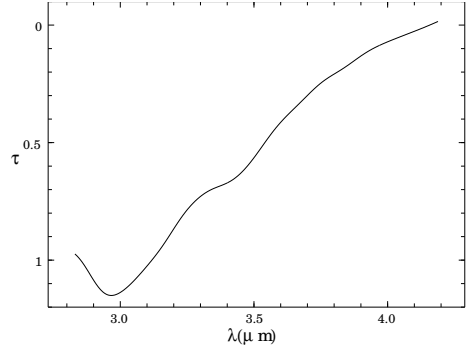}}} 
  \caption[]{Optical depth spectrum of the line of sight extinction 
    towards the Galactic Center.}
\label{Taulineofsight4.eps}
\end{figure}

The comparison of the corrected spectra from extinction (using the CO-star) and the blackbody continuum spectra with temperatures equal to the ones obtained by the fitting procedure (Sect. \ref{sec:Fitting} and Table~\ref{tabfit}) shows that the overall shape of the corrected spectra is consistent with the blackbody continuum shape of the corresponding balckbody temperature. 
The dusty sources are fitted well with blackbody temperatures of the order of $\sim 1000\,K$ and the global spectral shape of the hot
Helium stars in the IRS16 complex closely resembles pure Rayleigh Jeans
spectra. This implies that the L-band optical depth spectrum of
Fig. \ref{Taulineofsight4.eps} is consistent with the extinction law
of Martin \& Whittet (1990) (see section \ref{sec:Fitting}) and with
the known absorption value $2.7 \leq A_K \leq 4$ toward the Galactic
Center (Rieke \& Lebofsky 1985). Consequently, the fitting procedure made in Sect. \ref{sec:Fitting} is reliable and one can use the reddened continua obtained in that section in order to derive optical depth spectra. The extraction of the optical depth spectra is described in the next section. In the following, we do not make use of the corrected spectra obtained in Sect. \ref{sec:contr} because small absorption features could still be present in the spectrum of the correction CO-star which may affect the optical depth measurments. 

In addition, the good agreement between the two calibration procedures implies that
the overall variation in extinction across the central 0.5pc of the
Milky Way cannot be much larger than $\Delta$$A_K$=$\pm$0.5
magnitudes.  This is consistent with the results by Scoville et
al. (2003) who derived extinction estimates from the P$\alpha$/6cm
radio continuum and the P$\alpha$/H92$\alpha$ line emission over this
central region.  These estimates result in a distribution which is
smooth on the $\sim$1'' scale.  This supports the assumption that the
excess extinction seen towards some of the sources must in fact be
associated with the individual objects rather than with the diffuse
ISM (see Blum et al., 1996; Cl\'enet et al., 2001).

\section{Optical Depth Spectra}\label{sec:opticaldepth} 

\subsection{Extraction of Optical Depth Spectra}\label{sec:devtau}

As it has been pointed out in the previous section, the optical depth spectra of the Galactic Center sources can be obtained using the reddened blackbody continua of the fitting procedure (Sect. \ref{sec:Fitting} and Figs. \ref{Fitspectra1.eps} to \ref{Fitspectra3.eps}).\\

\begin{figure}
  \resizebox{8cm}{!}{\rotatebox{0}{\includegraphics{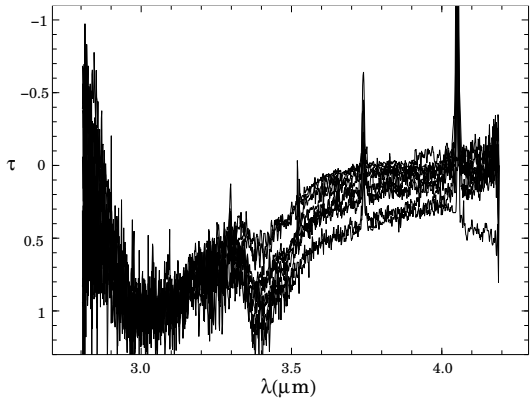}}}
  \caption[]{Optical depths of the Galactic Center sources 
          normalized to unity at $3.0\mu m$.}
\label{Tautous.eps}
\end{figure}

With the results of the blackbody fitting (Table \ref{tabfit}),
we have derived the optical depth spectra of the sources using the
equation $F_{obs}=F_{intr}e^{(-\tau)}$ where $\tau$ is the optical
depth, and $F_{obs}$ and F$_{intr}$ are the observed and intrinsic
fluxes, respectively.  All the spectra are shown in
Fig.~\ref{Tautous.eps} where they have been normalised to unity at
$3.0\mu m$.  Mean optical depth spectra for the dust enshrouded
sources and the non-obscured sources are shown in
Fig.~\ref{meanspectra-combi}.  This figure shows very clearly three
absorption features at $3.0\mu m$, $3.4\mu m$ and $3.48\mu m$.  From
Fig. \ref{Tautous.eps}, it is evident that the absorption depths of
the $3.4 \mu m$, $3.48 \mu m$ and $3.0 \mu m$ features vary from source to source.

\begin{figure*}
  \resizebox{15cm}{!}{{\includegraphics{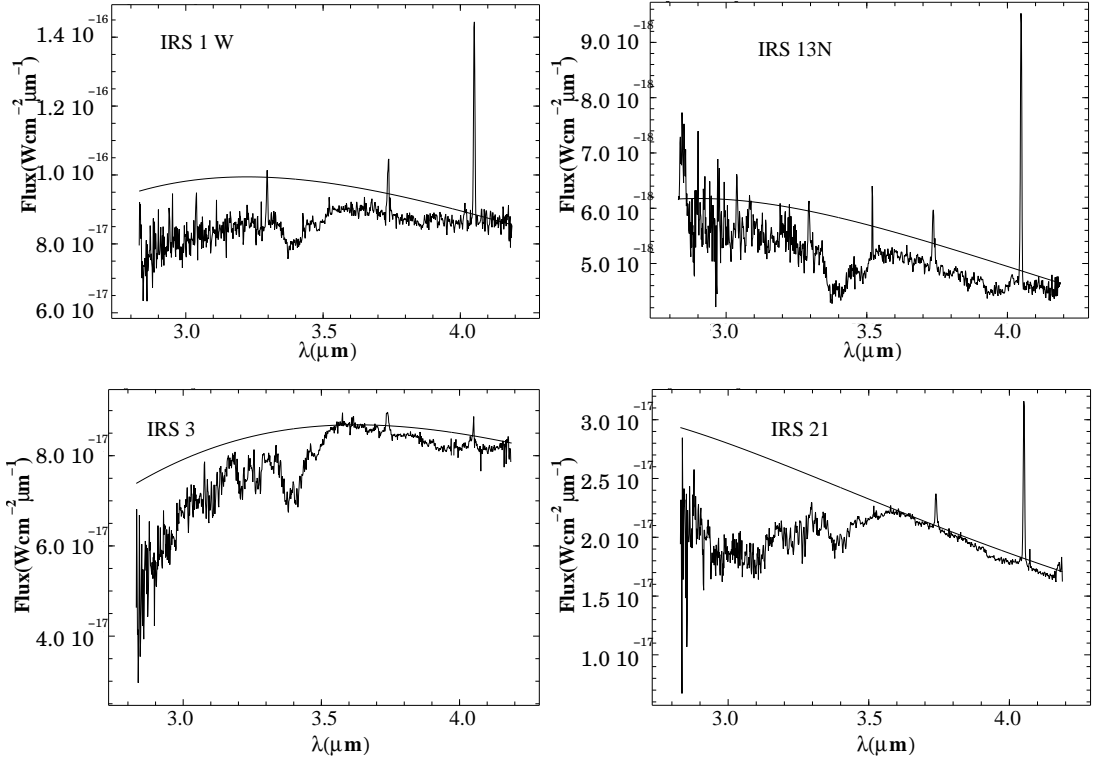}}}
  \caption[]{ L-band spectra corrected for line of sight absorption
using the optical depth spectrum shown in Fig. \ref{Taulineofsight4.eps}. 
Blackbody continua of temperatures listed in Table \ref{tabfit} 
are also plotted for comparison. Here no reddening of the blackbody continua 
is imployed.}
\label{Extcorr1.eps}
\end{figure*}

\begin{figure*}
  \resizebox{15cm}{!}{{\includegraphics{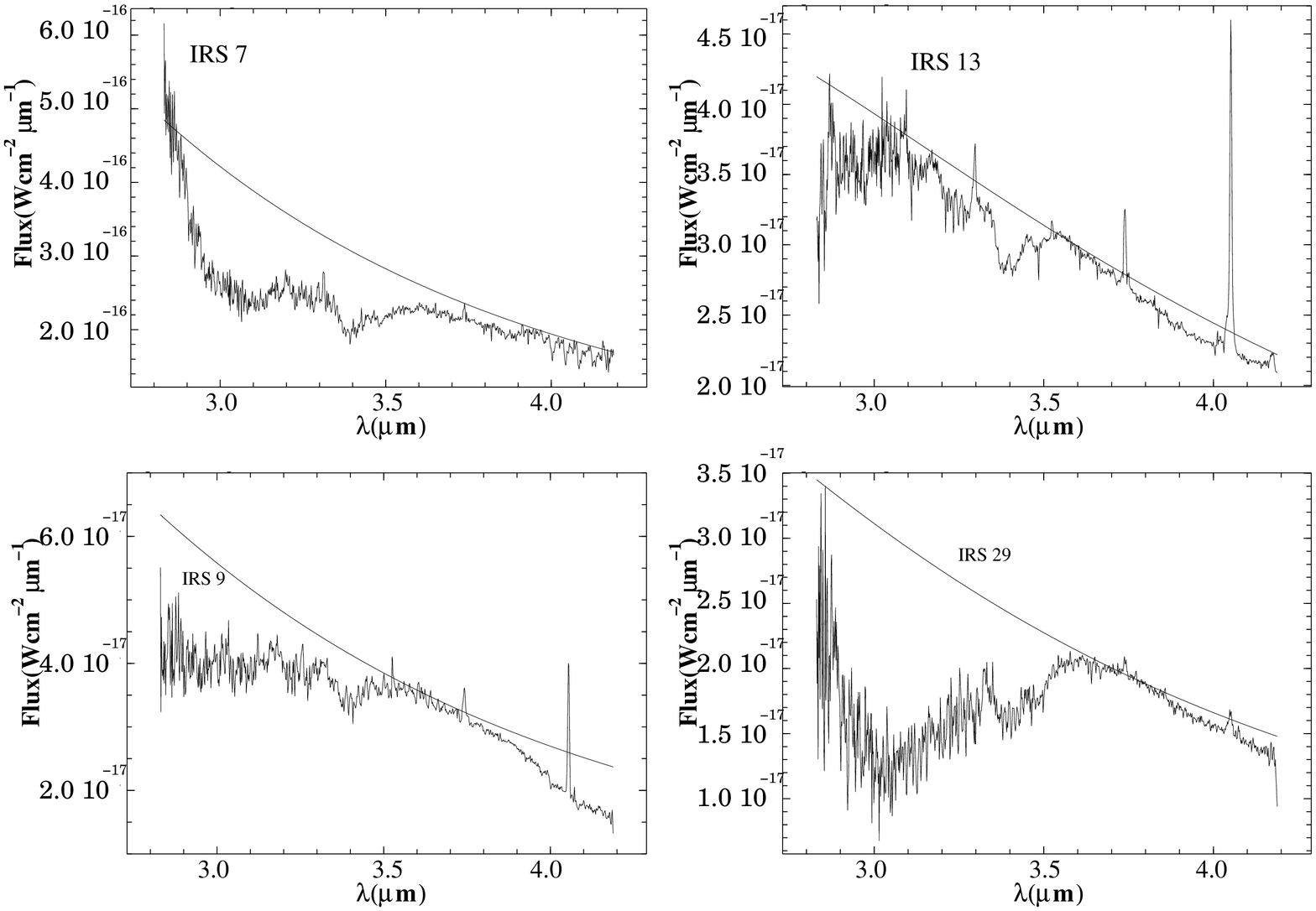}}}
  \caption[]{L-band spectra corrected for line of sight absorption (cont.).}
\label{Extcorr2.eps}
\end{figure*}

Mean spectra of the hydrocarbon feature at $3.4 \mu m$ and the ice
feature at $3.0\mu m$ have been constructed in the same iterative way
as in Chiar et al. (2002).  The optical depth spectrum of IRS~29 shows
the smallest $3.4\mu m$ absorption in comparison to the $3.0\mu m$
feature.  A first mean spectrum of the $3.4\mu m$ feature was obtained
by subtracting the optical depth spectrum of IRS~29 from all other
spectra.  Then the average spectrum of the residual spectra was
derived.  The mean water profile spectrum is then obtained by
subtracting the average spectrum of the $3.4\mu m$ hydrocarbon feature
multiplied by a free scaling parameter. This parameter is adjusted
such that a featureless spectrum is obtained in the $3.4\mu m$ region
after subtraction from the mean source spectrum.  Finally, the
spectrum of the $3.4\mu m$ hydrocarbon feature was deduced by
subtracting the mean $3.0\mu m$ spectrum from all the spectra and
averaging the residuals.  The resulting spectra are shown in Figs.
\ref{resid3-combi} and \ref{residu3.4-combi}.

The most reliable results were obtained for the dust enshrouded
sources.  As a consequence of the successful continuum fits described
in Sect. \ref{sec:Fitting}, the absorption spectra level out at $\tau$$\sim$0.0
for wavelengths well above and well below the corresponding absorption
features.  For the hot He-stars in the IRS16 cluster the situation is
different.  Here the continuum fits were less successful (see
discussion in Sect. \ref{sec:Fitting}) and the derived profile of the absorption
features are less reliable towards the long wavelength end of the
spectra. This is less severe for the $3.0\mu m$ feature since in this
case only the continuum long-ward of $\sim$3.5$\mu$m is affected.

\subsection{Characteristics of Features in the Optical Depth Spectra} \label{sec:structau}

\begin{figure}
  \resizebox{8cm}{!}{\rotatebox{0}{\includegraphics{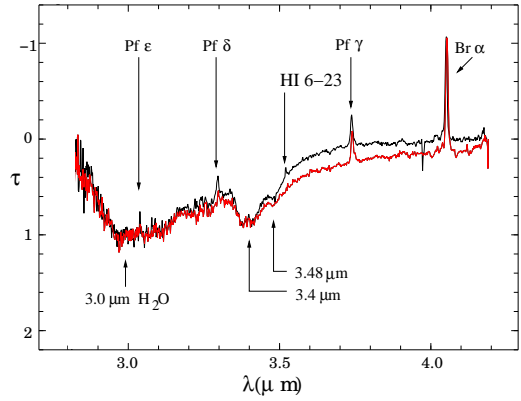}}}
  \caption[]{Mean L-band normalized spectra. 
The black and grey (or red in a colour version) spectra are obtained by averaging the dusty sources 
and the stellar (He-stars, IRS~7 and IRS~9) optical depth spectra, 
respectively.}
\label{meanspectra-combi}
\end{figure}

\begin{figure}
  \resizebox{8cm}{!}{\rotatebox{0}{\includegraphics{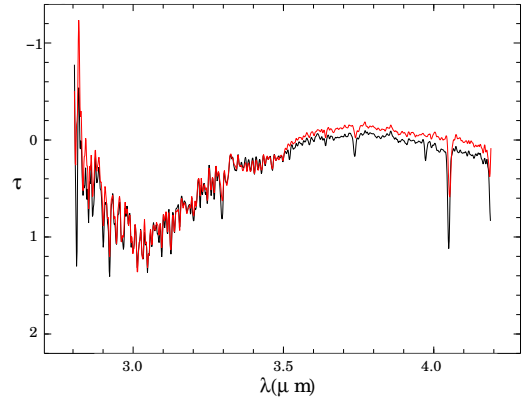}}}
  \caption[]{Mean isolated and normalized ice spectra of the dusty sources 
(in black) and He-stars, IRS~7 and IRS~9 sources (in grey or red in a colour version).}
\label{resid3-combi}
\end{figure}

\begin{figure}
  \resizebox{8cm}{!}{\rotatebox{0}{\includegraphics{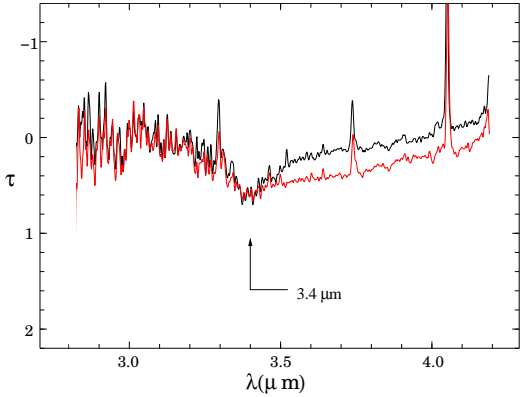}}} %{resid3.4norm.eps}}}
  \caption[]{Mean isolated and normalized $3.4\mu m$ feature of the dusty sources 
(in black) and He-stars, IRS~7 and IRS~9 sources (in grey or red in a colour version). 
The location of the $3.3\mu m$ feature is indicated.}
\label{residu3.4-combi}
\end{figure}

An examination of the optical depth spectra reveals information on the
origin and physical conditions of the absorbing material in the
central parsec of the GC stellar cluster. We distinguish the following
features:

\begin{itemize}
\item {\it The $3.0\mu m$ feature:} The dusty sources (which are located in the minispiral area like IRS~1W, 3, 13, 13N and 29) and hot He-stars (i.e. showing He emission lines in their K-band spectra which is the case for the IRS~16 sources, e.g. Najarro et al. 1997) allow to derive both the linewidth and line minimum of the 3.0$\mu$m
absorption feature.  Both quantities are not affected by the slight
continuum mismatch at wavelengths above $3.5\mathrm{\mu}$m in case of
the He-stars.  The $3.0\mu m$ absorption feature is associated with
interstellar $H_2O$ ice in molecular clouds (Gillett \& Forrest 1973).
Laboratory experiments suggest that the shape of the observed ice
absorption (which is deepest at about 2.95$\mu$m in IRS~7 for example)
toward the Galactic Center corresponds to trapped water ice in SiO
condensate (Wada et al. 1991).

Our mean absorption spectrum (Fig. \ref{resid3-combi}) shows,
however, that in our observations the deepest absorption occurs
$\sim 0.05\mathrm{\mu}$m longward of 3.0$\mu$m.  This behavior is
consistent with the presence of amorphous $H_2O$ ice (Wada et
al. 1991).

Our mean spectrum of the 3.0$\mu$m ice feature appears to be narrower
than the one obtained by Chiar et al. (2002).  As modeled in their
paper, the 3.0$\mu$m ice feature derived by the authors is best fitted
by a temperature of 10K and a maximum mantle thickness of 0.75 to
0.85$\mu m$ assuming the simple model of core-mantle grains of Bohren
\& Huffman (1983) and the hypothesis of a variable mantle thickness of
Smith et al. (1993).  
%&While the model could only fit the blue part of
%their spectrum (from 2.8 to 3.2 $\mu m$), the optical depth spectrum
%derived by us is from $2.8\mu m$ to $3.5\mu m$ very consistent with a
%model of 40K ice temperature and a maximum mantle thickness of 0.85
%$\mu m$. The resulting temperature of 40K is compatible with the more
%central location (radius $\le$2.5'') of the sources studied in this
%paper in contrast to the sources considered in the paper of Chiar et
%al. (2002) which are located at larger separations from the center
%(radius mostly $\ge$2.5'').

%\item {\it The $3.3\mu m$ feature:} In our investigation we have also
%considered the optical depth at the location of the weak $3.3 \mu m$
%absorption feature as described by Chiar et al. (2002).  In our data
%the feature is clearest in the mean normalized L-band spectrum (see
%Figs. \ref{meanspectra-combi} and \ref{residu3.4}) as a weak depression in the red wing
%of the 3$\mu$m H$_2$O absorption.

\item {\it The 3.4$\mu$m and 3.48$\mu$m features} are known to arise
in the diffuse interstellar medium. 
They are both part of a single line complex (Duley \& Williams 1983)
and have already been seen towards many Galactic Center sources
(Sandford et al. 1991, Pendleton et al. 1994, Chiar et al. 2002,
Mennella et al. 2003).
They are most likely due to aliphatic hydrocarbons which are
characterized by their CH$_2$ (methylene) and CH$_3$ (methyl)
stretching modes (Sandford et al. 1991, Sellgren et al. 1995, Brooke, Sellgren \& Geballe
1999, Grishko \& Duley 2002, Butchart et al. 1986, Duley \& Williams
1984).\\
Actually, the $3.4\mu$m feature is composed of two bands, one at $3.38 \mu$m and the other at $3.42 \mu$m. These are due to the {\em asymmetric} stretching vibrations of $CH_3$ and $CH_2$ groups respectively (Sandford et al. 1991). \\
The $3.48\mu$m seen in all spectra (see also the mean optical depth 
spectrum in Fig. \ref{meanspectra-combi}) is due to the perturbed {\em symmetric} C-H stretching vibrations of the same $CH_3$ and $CH_2$ groups (Sandford et al. 1991). This feature was already reported by 
Allen \& Wickramasinghe (1981), Jones et al. (1983) and 
Butchart et al. (1986). 

The $3.4\mu m$ absorption feature for the dusty sources and 
the He-stars are shown in Fig. \ref{residu3.4-combi}.
For both types of sources the short wavelength wing is very
similar. The long wavelength wing of the line feature derived from the
He-stars is affected by the continuum mismatch discussed in Sect. \ref{sec:Fitting}.

%The absorption wing appears to be softer than in the case of the
%$3.4\mu m$ optical depth spectrum derived by Chiar et al. (2002).

\end{itemize}

\begin{figure}
  \resizebox{8cm}{!}{\rotatebox{0}{\includegraphics{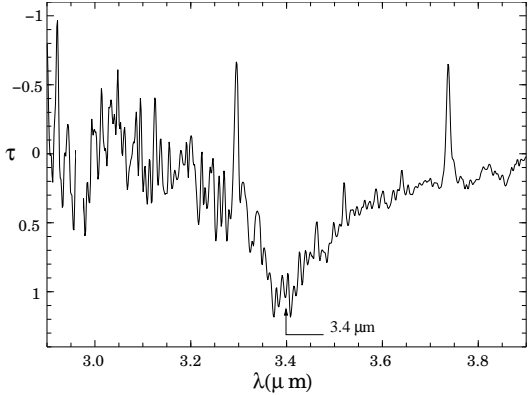}}}
  \caption[]{Mean isolated and normalized $3.4\mu m$ feature of the dusty sources. 
The location of the $3.3\mu m$ feature is indicated.}
\label{residu3.4}
\end{figure}

\begin{table*}[htbp]
\small
\begin{center}
\begin{tabular}{|l|c|c|c|c|}
\hline
Object & $\tau_{3.0\mu m}$ &  $\tau_{3.4\mu m}$ &  $\tau_{3.4\mu m}$ lower limit & $\tau_{3.48\mu m}$\\
\hline
IRS13  &$ 0.44\pm 0.03 $&$ 0.26\pm 0.02 $ &$ 0.16\pm 0.02 $ &$ 0.19\pm0.02$\\
IRS13N &$ 0.35\pm0.04 $&$ 0.33\pm 0.04 $ &$ 0.21\pm 0.03 $ &$ 0.25\pm 0.03 $\\
IRS16C &$ 0.67\pm 0.04 $&$ 0.49\pm 0.05 $ &$ 0.14\pm 0.02 $ &$ 0.45\pm0.03 $\\
IRS16CC &$ 0.425\pm 0.05 $&$ 0.36\pm 0.03$ &$ 0.16\pm 0.03 $ &$ 0.31\pm0.02$\\
IRS16NE &$ 0.38\pm 0.05 $&$ 0.27\pm 0.02 $ &$ 0.14\pm 0.02 $ &$ 0.23\pm0.02$\\
IRS16SW &$ 0.57\pm 0.03 $&$ 0.36\pm 0.05 $ &$ 0.12\pm 0.02 $ &$ 0.29\pm0.03$\\
IRS1W &$ 0.51\pm 0.28 $&$ 0.32\pm 0.04 $ &$ 0.16\pm 0.02 $ &$ 0.27\pm0.02$\\
IRS21 &$ 0.58\pm 0.03 $&$ 0.23\pm 0.04 $ &$ 0.15\pm 0.03 $ &$ 0.16\pm0.02$\\
IRS29 &$ 1.12\pm 0.09 $&$ 0.37\pm 0.03 $ &$ 0.27\pm 0.03 $ &$ 0.29\pm0.02$\\
IRS3 &$ 0.30\pm 0.02 $&$ 0.23\pm 0.03 $ &$ 0.20\pm 0.02 $ &$ 0.15\pm0.01$\\
IRS7 &$ 0.68\pm 0.03 $&$ 0.41\pm 0.06 $ &$ 0.20\pm 0.05 $ &$ 0.33\pm0.04$\\
IRS9 &$ 0.67\pm 0.06 $&$ 0.27\pm 0.07 $ &$ 0.16\pm 0.04 $ &$ 0.19\pm0.03$\\
\hline
\end{tabular}
\end{center}
\caption{The optical depths of the 4 absorption features at 
$3.0\mu m$, $3.3\mu m$, $3.4\mu m$ and  $3.48\mu m$.}
\label{taboptdept}
\end{table*}

%\begin{table*}[htbp]
%\small
%\begin{center}
%\begin{tabular}{|l|c|c|c|c|}
%\hline
%Object & $\tau_{3.0\mu m}$ & $\tau_{3.3\mu m}$ & $\tau_{3.4\mu m}$ & $\tau_{3.48\mu m}$\\
%\hline
%IRS13  &$ 0.44\pm 0.03 $&$  0.13\pm 0.07 $&$ 0.26\pm 0.02 $ &$ 0.19\pm0.02$\\
%IRS13N &$ 0.35\pm0.04 $&$ 0.18\pm 0.03 $&$ 0.33\pm 0.04 $ &$ 0.25\pm 0.03 $\\
%IRS16C &$ 0.67\pm 0.04 $&$ 0.32\pm 0.04 $&$ 0.49\pm 0.05 $ &$ 0.45\pm0.03 $\\
%IRS16CC &$ 0.425\pm 0.05 $&$ 0.21\pm 0.08 $&$ 0.36\pm 0.03$ &$ 0.31\pm0.02$\\
%IRS16NE &$ 0.38\pm 0.05 $&$ 0.17\pm 0.05 $&$ 0.27\pm 0.02 $ &$ 0.23\pm0.02$\\
%IRS16SW &$ 0.57\pm 0.03 $&$ 0.23\pm 0.05 $&$ 0.36\pm 0.05 $ &$ 0.29\pm0.03$\\
%IRS1W &$ 0.51\pm 0.01 $&$ 0.21\pm 0.04 $&$ 0.32\pm 0.04 $ &$ 0.27\pm0.02$\\
%IRS21 &$ 0.58\pm 0.03 $&$ 0.13\pm 0.03 $&$ 0.23\pm 0.04 $ &$ 0.16\pm0.02$\\
%IRS29 &$ 1.12\pm 0.09 $&$ 0.25\pm 0.03 $&$ 0.37\pm 0.03 $ &$ 0.29\pm0.02$\\
%IRS3 &$ 0.30\pm 0.02 $&$ 0.09\pm 0.04 $&$ 0.23\pm 0.03 $ &$ 0.15\pm0.01$\\
%IRS7 &$ 0.68\pm 0.03 $&$ 0.23\pm 0.04 $&$ 0.41\pm 0.06 $ &$ 0.33\pm0.04$\\
%IRS9 &$ 0.67\pm 0.06 $&$ 0.15\pm 0.09 $&$ 0.27\pm 0.07 $ &$ 0.19\pm0.03$\\
%\hline
%\end{tabular}
%\end{center}
%\caption{The optical depths of the 4 absorption features at 
%$3.0\mu m$, $3.3\mu m$, $3.4\mu m$ and  $3.48\mu m$.}
%\label{taboptdept}
%\end{table*}
\subsection{Measurements of the strengths of the Absorption Features}\label{sec:measurements}
The optical depths of the 4 absorption features at $3.0\mu m$, %$3.3\mum$,
 $3.4\mu m$ and $3.48\mu m$ are listed in Table~\ref{taboptdept}.
The values are obtained by averaging the optical depth spectra over
the wavelength intervals ]2.95$\mu m$,3.05$\mu m$[, %]3.30$\mum$,3.35$\mu m$[,
  ]3.35$\mu m$,3.45$\mu m$[ (including thus the two bands of the feature) and ]3.43$\mu m$,3.53$\mu
m$[, respectively.  The %$3.3\mu m$, 
 $3.4\mu m$ and $3.48\mu m$ optical
depths are measured in the spectra from which the $3\mu m$ ice
absorption feature was subtracted (see Sect. \ref{sec:devtau}).\\
The error bars listed in the table take only into account the measuring uncertainties, except for the $3.0\mu m$ feature of the IRS~1W case where the error bar is taken such that the optical depth value includes its upper and lower limits obtained by fitting upper and lower baselines as described in Sect.~\ref{fitresults}.\\
To estimate the lower limit of the optical depth values of the $3.4\mu m$ absorption feature, we measured the value of the strength of this feature using a linear continuum between $3.25\mu m$ and $3.65\mu m$ as baseline. The obtained values are listed in Table~\ref{taboptdept}.

Table \ref{taboptdept} shows that the optical depth values span a large interval suggesting that part of the absorption features arise probably from the local
 medium and may be associated with the individual sources.

Concerning the optical depth values of the $3.0\mu m$ feature, they agree well with the values obtained by Chiar et al. (2002) for the two sources in common IRS~1W and IRS~7. The $3.4\mu m$ feature of these two sources exhibits higher values than those of Chiar et al. (2002). This is probably the result of the derivation of the
isolated ice absorption feature described in Sect. \ref{sec:devtau}
which is sensitive to the choice of the subtracted spectrum (here the
spectrum of IRS~29). In particular, in Chiar et al. (2002), the
authors find a negative optical depth value at $3.3\mu m$ for IRS~7
due to the shape of the ice feature which is narrower than the mean
ice absorption feature. Moreover, their optical depth value of the $3.4\mu m$ feature of IRS~7 is even smaller than the lower limit obtained in our spectrum.\\
Concerning the IRS~3 source results, no comparison can be done with Chiar et al. (2002) as the fit of their spectrum was not satisfactory and did not match at all the K-band point flux. This may be due to the problematic relative flux calibration of their L-band spectrum which shows a different slope than ours.

On the other hand, the lower limits of the optical depth values of the hydrocarbon feature for the IRS~3 and IRS~7 sources in common with Sandford et al. (1991, 1995) and Pendelton et al. (1994) agree well with those obtained by these authors. This is very satifactory as the values provided in these papers were derived using a linear fit similar to the one used here to derive the lower limits.\\

\subsection{Correlations Between the Strengths of the Absorption Features} \label{sec:corr}

\begin{figure}
  \resizebox{7cm}{!}{\rotatebox{0}{\includegraphics{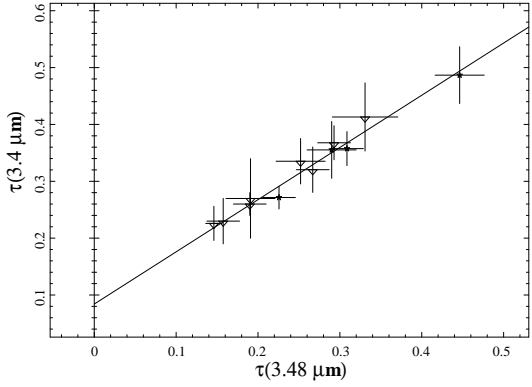}}}
  \caption[]{Optical depth of the hydrocarbon absorption  
   at the $3.4\mu m$ versus the $3.48\mu m$ absorption feature.}
\label{figtau3.4v3.48}
\end{figure}

\begin{figure}
  \resizebox{7cm}{!}{\rotatebox{0}{\includegraphics{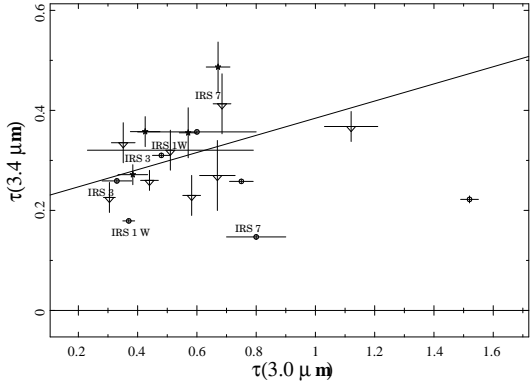}}}
  \caption[]{Optical depth of the hydrocarbon absorption at $3.4\mu m$ 
   versus the depth in the $3.0\mu m$ absorption feature 
   for all observed Galactic Center sources (the IRS~16 sources are represented by stars and the remaining sources by triangles). The values obtained by Chiar et al. (2002) are also shown (in circles) but are not considered in the linear regression fit.}
\label{figtau3.4v3}
\end{figure}

%Concerning the $3.3\mu m$ absorption feature, we have averaged the
%flux over the red part of the feature in order to avoid a bias due to
%the presence of the Pf$\delta$ hydrogen emission line at $3.296\mu m$
%in some of the spectra.  The large uncertainties found for the $3.3\mu
%m$ feature are likely due to the contamination of this absorption
%feature by the hydrogen emission line. 
The plots of the optical
depths of the absorption features at %$3.0\mu m$ versus $3.3\mu m$,
 $3.4\mu m$ versus $3.0\mu m$ %, $3.4\mu m$ versus $3.3\mu m$, 
 and $3.4\mu m$ versus $3.48\mu m$ are shown in Figs. %\ref{figtau3v3.3},
\ref{figtau3.4v3} %, \ref{figtau3.4v3.3}, 
 and \ref{figtau3.4v3.48} respectively. In these figures we also plot the data
 by Chiar et al. (2001) that were mostly obtained on sources in the outer parts of the central stellar cluster.

We find a good correlation (correlation coefficient of $\sim 0.98$) between the $3.4\mu m$ and $3.48\mu m$ absorption features (Fig. \ref{figtau3.4v3.48}) and a trend of correlation (correlation coefficient of $\sim 0.43$) between the $3.4\mu m$ and $3.0\mu m$ features (Fig. \ref{figtau3.4v3}). In Figs. \ref{figtau3.4v3.48} and \ref{figtau3.4v3}, the best fits of linear regression are drawn as well; the goodness of fit probabillities are respectively $0.9996$ and $8\,10^{-4}$. The first correlation is forseeable since both features (at $3.4\mu m$ and $3.48\mu m$) arise from the same functional groups as explained in Sect. \ref{sec:structau}. %This result provides an additional evidence that the reddened blackbody baselines obtained by the fitting procedure in Sect. \ref{sec:Fitting} are reliable, because otherwise, one does not expect both features to be correlated.

The possible correlation between the optical depths of
the $3.4\mu m$ and $3.0\mu m$ features shown in Fig.
\ref{figtau3.4v3} suggests that the ISM in the central region is a
mixture of diffuse and dense material. On the other hand, if this correlation is real, then in this plot the
discrepancy between the positions of IRS~7 derived by us and by Chiar
et al. (2002) cannot only be due to the variability of the source. In
the case of variability, both positions should follow the overall
trend of the correlation. Actually, the discrepancy between the
positions of the 3 sources (IRS~1W, IRS~3 and IRS~7) in common with
Chiar et al. (2002) is due to the same reasons evoked previously in Sect. \ref{sec:measurements}. 

%We find correlations between all 4 absorption line
%features.  

%In the $3.0\mu m$ versus $3.3\mu m$ plot in Fig. \ref{figtau3v3.3},
%our data show a tendency for a correlation that was not contained in
%the results of Chiar et al. (2002).  Such a correlation would be
%expected if indeed a substantial fraction of the $3.3\mu m$ feature was
%directly associated with the individual sources or the local ISM in
%the central parsec.

%Similar to the work by Chiar et al. (2001) our data are consistent
%with a correlation between the $3.4\mu m$ hydrocarbon and the $3.3\mu
%m$ features (see Fig. \ref{figtau3.4v3.3}).  This also supports the
%assumption that this feature is connected to the diffuse ISM in the
%very center of the Galaxy.

%We also find a good correlation between the 3.4$\mu$m and 3.48$\mu$m
%absorption features (Fig. \ref{figtau3.4v3.48}).  

The fact that the
$3.4\mu m$ to $3.0\mu m$ %(better seen in the $3.4\mu m$ to $3.3\mu m$
%correlation which we also show for clarity) 
% and the $3.4\mu m$ to
%$3.48\mu m$ 
absorption feature ``correlation'' shows a significant offset
from the origin of the coordinate system used for display may indicate
that there could be a significant $3.4\mu m$ line absorption even if the
water ice feature were not present. Therefore we  may conclude that
a certain amount %approximately 50\% (in $\tau$ and therefore in the corresponding column density) 
of the 3.4$\mu$m line absorption is due to the diffuse
ISM on the line of sight to the Galactic Center. Due to the low density of the
mini-spiral gas a substantial amount may, in fact, be closely linked
to the individual sources that we are studying here in detail.
Therefore, the remaining portion of the $3.4\mu m$ absorption may occur in the central parsec. This is also supported by the finding of Sandford et al. (1995) who derived anomalously high aliphatic CH absorption per visual extinction relative to all the other lines of sight for which data was then available.

\begin{figure}
\begin{center}
\includegraphics[width=8cm, height=6cm, angle=0]{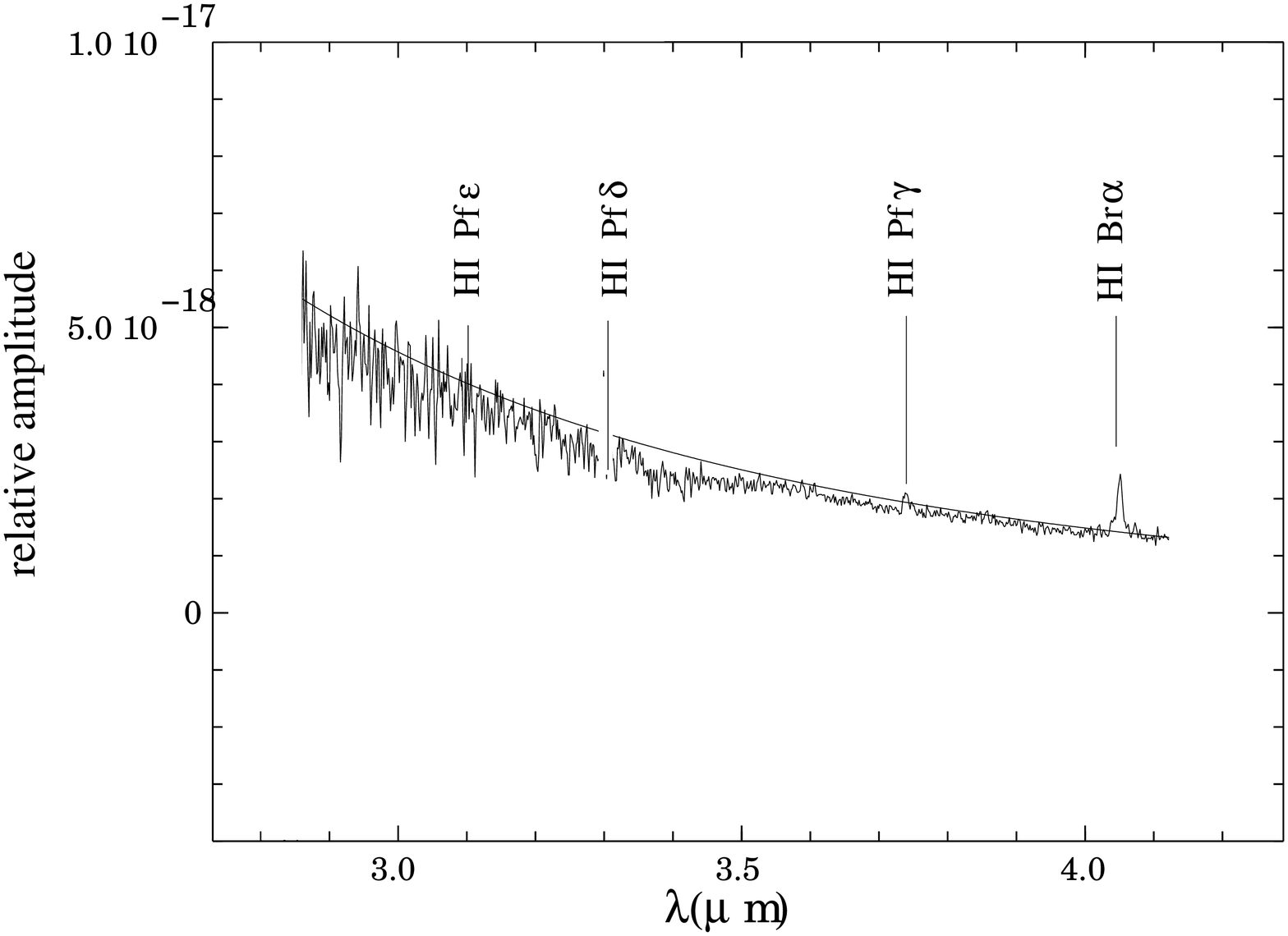}
\caption{L-band intrinsic spectrum (i.e. corrected from extinction along the line of sight as explained in Sect. \ref{sec:contr}) at the position of SgrA* taken through a 0.6`` wide
slit using ISAAC at the VLT UT1. The continuum of a blackbody of 22,200~K 
is also shown.  The locations of hydrogen lines are marked. }
\label{eckart5.eps}
\end{center}
\end{figure}

\subsection{Hydrogen emission lines} \label{sec:Hlines}

The obtained L-band spectra also allow us to investigate the HI line emission 
towards selected lines of sight. 
The corresponding line strengths, widths and equivalent widths for
Br$\alpha$, Pf$\gamma$, and Pf$\delta$ are listed
in Tables \ref{HItab1}, \ref{HItab2}, and \ref{HItab3}.
At the achieved spectral resolution we cannot distinguish 
between the emission from
the individual objects and the mini-spiral. 
The latter seems to dominate the emission
since sources on or close to the mini-spiral 
(IRS~1W, 16NE, 16SW, 21, 13, 13N), 
also show the strongest line emission.
Especially for IRS~21 the K-band spectrum by Ott et al. (1999) 
suggests that probably all
the hydrogen line emission is due to the mini-spiral.
\\
Using the $Br_{\gamma}$ and $P_{\alpha}$ line strengths published by 
Najarro et al. (1997) we can calculate the diagnostic line ratios 
$Br_{\gamma}/Br_{\alpha}$ and $P_{\alpha}/Br_{\alpha}$  
for three of the IRS~16 sources.
The results are shown in Table \ref{tabratios}. 
In order to obtain dereddened line strengths, we have considered an absorption 
in the L-band wavelength range of $A_L=1.68$ consistent with our optical 
depth spectrum of Fig. \ref{Taulineofsight4.eps} as well as the value given 
by Rieke \& Lebofsky (1985).
\\
Taking the dereddened ratios at face value and omitting possible small
contributions from the He line emission close to P$\alpha$ and Br$\gamma$
line we can derive first estimates about the temperature and density 
of the emitting gas.
Assuming case B recombination, for both sources IRS~16NE and IRS~16C
the line ratios are consistent with emission from gas at low temperatures
of $\le 5000$K and low densities of $\le$10$^2$cm$^{-3}$.
For IRS~16SW the ratios are more consistent with  emission from a gas at
high temperature $\sim 20,000$K and high density $\sim 10^4$cm$^{-3}$.
While it cannot be excluded that the emission towards IRS~16SW contains a 
significant contribution 
from the hot stellar atmospheres of the He-star IRS~16SW and the other
neighboring hot stars in the IRS~16 complex,
this result reflects the complex density and temperature structure 
within the mini-spiral that consists of a thermal plasma of
n$_e$$\sim$10$^4$cm$^{-3}$ at T$\sim$10$^4$~K
(e.g. Brown, Johnston, \& Lo 1981)
with denser entities of n$_H$$>$10$^5$cm$^{-3}$ and a few 100~K
(e.g. Jackson et al. 1993).
 
\begin{table*}[htbp]
\small
\begin{center}
\begin{tabular}{|l|c|c|}
\hline
Object & $(P_{\alpha}+He)/Br_{\alpha}$ & $(Br_{\gamma}+He)/Br_{\alpha}$ \\
\hline
IRS16NE &$ 3.2 \pm 0.6 $&$ 0.28 \pm 0.03 $\\   
IRS16C  &$ 3.4 \pm 0.6 $&$ 0.26 \pm 0.03 $\\   
IRS16SW &$ 4.6 \pm 0.8 $&$ 0.43 \pm 0.04 $\\   
\hline
\end{tabular}
\end{center}
\caption{Diagnostic line ratios for the three IRS~16 sources in common with Najarro et al. (1997).}
\label{tabratios}
\end{table*}

\subsection{L-band spectrum at the position of SgrA*} \label{sec:sgra}

Our ISAAC L-band imaging and spectroscopy was part of a
flux density monitoring program of SgrA* (Baganoff et al. 2001, 2003
and  Eckart et al. 2004)
which has been carried out simultaneously to CHANDRA observations.
Therefore all slit settings were defined such that the position
towards SgrA* always fell into the slit.
A mean extinction corrected spectrum at that position is shown in 
Fig. \ref{eckart5.eps}.
Here the weak hydrogen recombination lines
are most likely due to emission from the mini-spiral.
The Rayleigh Jeans shape of the continuum spectrum compares 
well to that of other hot stars, like the He stars in the central cluster. 
In Fig. \ref{eckart5.eps} we show the L-band spectrum of SgrA* 
compared to a 22,200K blackbody spectrum.
In fact, high resolution adaptive optics L-band images were taken
in August 2002 with NAOS/CONICA on the VLT UT4 during the science
verification phase (Cl\'enet et al. 2003, Genzel et al. 2003, Eckart et al. 2004).These images show that at the position of SgrA* the L-band flux
density is dominated by the fast moving star S2 (Sch\"odel et al. 2003).
Our spectrum is consistent with the fact that based on its K-band luminosity
spectrum  S2 is likely to be a $\sim$15-20 solar mass late O, early B
main-sequence star of age less than 20 Myr
(Gezari et al., 2002; Eckart et al., 1999; Figer et al., 2001; Ghez et
al., 2003). 

%\begin{figure}
%  \resizebox{7cm}{!}{\rotatebox{0}{\includegraphics{figTau3v3.3.eps}}}
%  \caption[]{Water ice optical depth at $3.0\mu m$ versus $3.3\mu m$ absorption 
%feature for all observed Galactic Center sources (triangles) and for 
%the sources of Chiar et al. (2002) (circles).}
%\label{figtau3v3.3}
%\end{figure}

%\begin{figure}
%  \resizebox{7cm}{!}{\rotatebox{0}{\includegraphics{figTau3.4v3.3.eps}}}
%  \caption[]{Hydrocarbon absorption feature at $3.4\mu m$ versus $3.3\mu m$ 
%   absorption feature for all observed Galactic Center sources (triangles) 
%   and for the sources of Chiar et al. (2002) (circles).}
%\label{figtau3.4v3.3}
%\end{figure}

\begin{table*}[htbp]
\small
\begin{center}
\begin{tabular}{|l|c|c|c|c|}
\hline
Object          & Line-Continuum of HI (4.051$\mu m$) & Integral of HI (4.051$\mu m$) & EW of HI (4.051$\mu m$) & Line width\\ 
                &  $ 10^{-18}\, W cm^{-2}\mu m{-1}$ & $10^{-18}\, W cm^{-2}$ & \AA\ & \AA\ \\
\hline
IRS 1 W &$55.9\pm0.40 $&$ 0.5869\pm0.1216  $&$ -41.48\pm1.86 $ & $210\pm42$\\  
IRS 3 &$5.80\pm0.50 $&$  0.0855\pm0.0378   $&$ -6.02\pm2.83 $ & $295\pm105$\\  
IRS 7 &$ -                    $&$ -                        $&$ -             $ & -\\  
IRS 9 &$19.60\pm0.40 $&$ 0.1137\pm0.0121  $&$ -63.56\pm1.64 $ & $116\pm10$\\  
IRS 21 &$13.00\pm0.20 $&$ 0.1566\pm0.0362  $&$ -48.03\pm1.90 $ & $241\pm52$\\ 
IRS 29 &$1.70\pm0.20 $&$  0.0196\pm0.0041 $&$ -10.26\pm1.75 $ & $231\pm21$\\  
IRS 13 &$47.50\pm0.40 $&$  0.8479\pm0.0570  $&$ -87.59\pm2.63$ & $357\pm21$\\ 
IRS 13 N    &$35.60\pm0.30 $&$ 0.2243\pm0.0393  $&$ -63.58\pm1.18 $ & $126\pm21$\\  
IRS 16 CC  &$1.69\pm0.03 $&$ 0.0275\pm0.0066  $&$ -58.20\pm4.39 $ & $326\pm73$\\  
IRS 16 C  &$3.86\pm0.06 $&$ 0.0588\pm0.0028  $&$ -117.43\pm3.28 $ & $305\pm10$\\  
IRS 16 NE   &$10.19\pm0.08$&$ 0.1126\pm0.0065$&$ -112.82\pm2.94 $ & $221\pm11$\\
IRS 16 SW &$6.29\pm0.07 $&$ 0.1025\pm0.0109 $&$ -120.66\pm3.61 $ & $326\pm31$\\\hline
\end{tabular}
\end{center}
\caption{Continuum and emission line strengths as well as linewidths 
and equivalent widths of the Br$\alpha$ line towards the observed sources.}
\label{HItab1}
\end{table*}

\begin{table*}[htbp]
\small
\begin{center}
\begin{tabular}{|l|c|c|c|c|}
\hline
Object          &  Line-Continuum of HI (3.739$\mu m$) & Integral of HI (3.739$\mu m$)& EW of HI (3.739$\mu m$) & Line width\\
                &  $ 10^{-18}\, W cm^{-2}\mu m{-1}$ & $ 10^{-18}\, W cm^{-2}$ & \AA\ & \AA\ \\
\hline
IRS 1 W     &$13.50\pm0.60$&$ 0.1276\pm0.0198$&$ -12.86\pm1.36 $ & $189\pm21$\\IRS 3       &$3.30\pm0.40$&$  0.0346\pm 0.0077$&$ -4.79\pm1.06$ & $210\pm21$\\ 
IRS 7       &$18.00\pm1.00$&$  0.1224\pm 0.0158$&$ -4.41\pm0.52 $ & $136\pm10$\\  
IRS 9  &$3.80\pm0.70$&$  0.0380\pm0.0091$&$ -12.63\pm1.67 $ & $200\pm11$\\  
IRS 21 &$2.10\pm0.10$&$  0.0166\pm0.0041$&$ -9.29\pm0.69 $ & $158\pm32$\\ 
IRS 29 &$1.10\pm0.10$&$  0.0072\pm0.0009$&$ -3.73\pm0.31 $ & $131\pm5$\\ 
IRS 13       &$8.90\pm0.25$&$ 0.0703\pm0.0064$&$ -15.29\pm0.86$ & $158\pm10$\\
IRS 13 N     &$6.30\pm0.30$&$ 0.0627\pm0.0127$&$ -15.38\pm1.64 $ & $199\pm31$\\
IRS 16 CC  &$0.37\pm0.03$&$ 0.0039\pm0.0007$&$ -9.96\pm1.14 $ & $210\pm21$\\
IRS 16 C  &$0.86\pm0.05$&$ 0.0086\pm0.0009$&$ -20.03\pm1.84 $ & $200\pm10$\\
IRS 16 NE   &$2.16\pm0.08$&$0.0239\pm0.0042$&$ -21.74\pm2.50 $ & $221\pm31$\\
IRS 16 SW  &$1.07\pm0.05$&$ 0.0096\pm0.0010$&$ -19.72\pm1.53 $ & $179\pm10$\\
\hline
\end{tabular}
\end{center}
\caption{Continuum and emission line strengths as well as linewidths 
and equivalent widths of the Pf$\gamma$ line towards the observed sources.}
\label{HItab2}
\label{measure1}
\end{table*}

\begin{table*}[htbp]
\small
\begin{center}
\begin{tabular}{|l|c|c|c|c|}
\hline
Object          & Line-Continuum of HI (3.296$\mu m$) & Integral of HI (3.296$\mu m$) & EW of HI (3.296$\mu m$) & Line width \\
                &  $ 10^{-18}\, W cm^{-2}\mu m{-1}$ & $ 10^{-18}\, W cm^{-2}$ & \AA\ & \AA\ \\
\hline
IRS 1 W     &$ 7.80\pm0.20    $&$ 0.0573\pm0.0097$&$ -9.26\pm0.81 $ & $147\pm21$\\
IRS 3       &$ -                        $&$ -                     $&$ -           $ & -\\  
IRS 7       &$ 12.00\pm2.00    $&$ 0.0450\pm0.0135$&$ -3.50\pm0.94 $ & $75\pm10$\\  
IRS 9  &$ -                        $&$ -                     $&$ -             $ & -\\  
IRS 21 &$ -                        $&$ -                     $&$ -             $ & -\\ 
IRS 29 &$ -                        $&$ -                     $&$ -             $ & -\\
IRS 13       &$4.80\pm0.20      $&$0.0226\pm0.0033$&$-7.65\pm0.68 $ & $94\pm10$\\   
IRS 13 N     &$ 3.30\pm0.30    $&$ 0.0208\pm0.0054$&$ -8.29\pm0.99 $ & $126\pm21$\\  
IRS 16 CC  &$ -                        $&$ -                     $&$ -             $ & -\\  
IRS 16 C  &$ 0.57\pm0.12    $&$0.0027\pm0.0008$&$ -7.57\pm1.19 $ & $94\pm10$\\  
IRS 16 NE   &$ 1.15\pm0.16    $&$0.0084\pm0.0024$&$ -7.62\pm1.31 $ & $147\pm21$\\  
IRS 16 SW  &$ 0.38\pm0.09    $&$0.0026\pm0.0008$&$ -4.30\pm1.18 $ & $137\pm10$\\  
\hline
\end{tabular}
\end{center}
\caption{Continuum and emission line strengths as well as linewidths and equivalent widths of the Pf$\delta$ line towards the observed sources.}
\label{HItab3}
\end{table*}

\section{IRS~13N} \label{sec:irs13n}

Well within the central stellar cluster of the Milky Way, about 0.5''
north of the IRS~13 complex,trend in Eckart et al. (2004) have found a
small cluster ($<$0.13 light years) of compact sources with a strong
infrared excess and obtained a first spectrum of it.  The integrated
spectrum of that area clearly shows that this excess is due to the
contribution of warm T=500~-~1000~K dust.  The nature of these newly
found sources is currently unclear.  Eckart et al. (2004) discuss
three possible explanations.  The L-band excess sources north of the
IRS~13 complex could be heavily extincted luminous stars. They could
also be hot stars that heat and interact with the more ambient 
environment of the local mini-spiral.  
Finally these newly found objects may even be young
stars and their L-band excess may be due to the flux density
contribution of luminous accretion disks.  While a combination of the
first two possibilities currently cannot be entirely excluded, the
luminosities and colors of these sources are consistent with those of
young stellar objects.  As described in Sect. \ref{sec:Fitting}, a
reddened blackbody continuum was also fitted to the IRS~13N spectrum
considering the K- and L-band fluxes given by Eckart et al. (2004) (see
Figs. \ref{Fitspectra2.eps} and \ref{Extcorr2.eps} and Table
\ref{tabfit}).  The resulting temperature and K-band extinction agree
well with the above explanations. The extinction of about $3.9$~mag in
the K-band is by about $1mag$ higher than the overall line of sight
extinction toward the Galactic Center and a $\sim$1000$K$ dust
temperature emission agrees well with the highly dust embedded YSO's
as studied by e.g. Ishii et al. (1998).  The ice absorption feature
often observed towards YSO's (see Ishii et al 1998) is not detectable
in the extinction corrected spectrum shown in Fig.
\ref{Extcorr1.eps}; this may in part be due to the small absorption in
the spectrum of the late-type CO-star that reverberates on the
spectrum of the extinction along the line of sight (Fig.
\ref{Taulineofsight4.eps}).  It is also possible that the
circumstellar material is affected by the intense radiation field at
the Galactic Center (Lutz et al. 1996).  In this case the ices in the
disks of young stars may become either tenuous and/or destroyed by
hard-UV photons.

\section{Summary and Conclusions} \label{sec:summary}
Combined NIR and MIR spectroscopic observations of sources in the
central $0.5$~pc parsec of the Milky Way allowed us to obtain a
detailed picture of the absorption features visible in the spectra
towards that region. Our investigation on the central sources (radius
$\le$2.5'') complements the study by Chiar et al. (2002) that mostly
includes sources with larger separations from the center. We find some
evidence that the diffuse ISM in the $0.5$~pc has properties that are
slightly distinct from the ISM at larger distances from the center.

The 3.0$\mu$m ice profile usually observed toward the Galactic Center
 peaks at 2.96$\mu$m, short-ward of the ice feature in local molecular
 clouds (McFadzean et al. 1989; Tielens et al. 1996; Chiar et
 al. 2000).  It is likely due to cold (15 K) water ices with an
 enhanced NH$_3$ abundance (Chiar et al. 2000).  Especially processed
 ices contribute a substantial portion of the refractory grain
 materials that persist when the molecular cloud is dispersed by star
 formation, and these products may yield the extinction characteristic
 of the diffuse interstellar medium.  

 However, the profile of the $3.0\mu m$ ice feature obtained in our work peaks
 long-ward of $3.0\mu m$ and therefore is likely to be associated with
 the presence of amorphous $H_2O$ ice towards these sources.  

Simultaneous fits of our K- and L-band spectra with single reddened
blackbody continua allowed us to estimate the extinction towards the
individual Galactic Center sources and to determine the approximate
continuum shape of the observed spectra.  The derived K-band
extinctions and blackbody temperatures are consistent with values
found for the sources in the central stellar cluster (Chiar et
al. 2002, Tanner et al. 2002, 2003).

Using the spectrum of a late-type star assumed to be free from local
extinction, we were able to derive the spectrum of the L-band
extinction along the line of sight toward the Galactic Center.  This
spectrum has been used in order to derive the extinction corrected
spectra of the most luminous sources.  The extinction corrected
spectra are consistent with the blackbody temperatures derived from
the previous fitting procedure.  The Rayleigh Jeans continuum spectra
obtained towards all the hot stars indicate that the distribution of
extinction in the central half parsec is fairly flat and varies in the
K-band only by $\Delta$A$_K$$\sim \pm 0.5$ mag.  This is consistent
with the results by Scoville et al. (2003). Therefore the excess
extinction that we determined after correction for the line of sight
extinction towards some of the sources should be associated with the
individual objects or very clumpy features of the ISM. This is in
support of the findings by Blum et al. (1996) and Cl\'enet et
al. (2001) and is also consistent with the large interval in which range the optical depth values derived from the fitting procedure.

The presence of local extinction in the envelopes of the dusty sources
is consistent with the bow shock model of Tanner et al. (2002, 2003).
Cotera et al. (1999) had already shown that several of these sources
are indeed offset from nearby local maxima in the extended dust
emission and temperature distribution.  Especially for IRS~21, Tanner
et al. (2002) indicate that the extended dust emission of this source
is consistent with a bow shock created by the motion of such a massive
hot star through the dust and gas of the mini-spiral.  It is likely
that the bow shock scenario may be applicable to most of the dust
embedded sources in the central stellar cluster. 
For IRS~3, however, the bow-shock scenario may not apply.
Gezari et al. (1985) find IRS~3 as the most compact and
(together with IRS~7) hottest, bright source (T$\sim$400~K) 
in the central cluster.
As suggested in the case of IRS~21 by Tanner et al. (2002)
IRS~3 may be an optically thick dust shell surrounding a
mass-losing source, such as a dusty recently formed WC9 Wolf-Rayet
star. 

In addition, a blackbody temperature of $\sim$ 1000K is found for the
spectrum representing the highly reddened sources located in the North
of IRS~13.  While it cannot be excluded that the individual objects
contained in this source complex are lower luminosity analogues of the
class of bow shock objects found by Tanner et al. (2002) and Rigaut et
al. (2003), their temperature and luminosity is well in agreement with
the low temperatures of the YSO classified by Ishii et al. (1998).

Detailed modeling, similar to the studies by Tanner et al. (2003) and
Eckart et al. (2004), based on higher angular resolution MIR imaging
and spectroscopy (using AO or interferometry) is required to unravel
the nature of the highly extinced sources in the IRS ~13N association.

Finally, a trend of correlation is noticed in the $3.4\mu m$ versus $3.0\mu m$ optical depths plot. If a real correlation were confirmed between these two features, it would suggest that the ISM along the line of sight toward the Galactic Center is possibly composed of a mixture of diffuse and dense material. Moreover, the plots of the optical depth values suggests that part
%about $50\%$ 
of the $3.4\mu m$ feature arises probably from the foreground ISM and %the remaining $50\%$ 
part of it from the local medium associated with the individual sources.

\bigskip

%===================================================================
\appendix

%============================================================

\begin{acknowledgement}
This work was supported in part by the Deutsche Forschungsgemeinschaft 
(DFG) via grant SFB 494.
We are grateful to all members of the ISAAC/VLT and the MPE 3D team.
\end{acknowledgement}

%\begin{references}
\bibliographystyle{apj}
%\bibliography{apj-jour,arraydesign,interfero,deconvolution,mosaicing}

\rf{Allen, D.A. \& Wickramasinghe, D.T. 1981, Nature 294, 239}

\rf{Baganoff, F. K., Bautz, M. W., Brandt, W. N., Chartas, G., 
  Feigelson, E. D., Garmire, G. P., Maeda, Y., Morris, M., 
  Ricker, G. R., Townsley, L. K., Walter, F. 2001, Nature, Volume 413, 
  Issue 6851, pp. 45-48 }

\rf{Baganoff et al.,
   Proceedings of the Galactic Center Workshop, Nov. 3-8, 2002, Hawaii,
   A. Cotera, T. Geballe, S. Markoff, H. Falcke (editors) 2003,
   Astron. Nachrichten in press }

\rf{Becklin, E.E. \& Neugebauer, G. 1968, iihconf 1 }

\rf{Becklin, E.E. \& Neugebauer, G. 1969, ApJ 157, L31 }

\rf{Becklin, E.E. \& Neugebauer, G. 1975, ApJ 200, L71 }

\rf{Becklin, E.E., Matthews, K., Neugebauer, G., Willner, S.P. 1978, ApJ 219, 121 }

\rf{Blum, R. D., Sellgren, K., Depoy, D. L. 1988, AJ 112}

\rf{Blum, R. D., Sellgren, K., Depoy, D. L. 1995, ApJL 440, L17 }

\rf{Blum, R. D., Depoy, D. L., Sellgren, K. 1995, ApJ 441, 603 }

\rf{Blum, R. D., Sellgren, K., Depoy, D. L. 1996, ApJ 470, 864 }

\rf{Bohren, C.F., Huffman, D.R. 1983, asls. book }

\rf{Brandner, W. et al. 2002, The ESO Messenger 107, 1-6 }

\rf{Brooke, T. Y., Sellgren, K., Geballe, T. R. 1999, ApJ 517, 883 }

\rf{Brooke, T. Y., Tokunaga, A. T., Strom, S. E. 1993, AJ 106, 656 }

\rf{Brown, R.L., Johnston, K.J., Lo, K.Y. 1981, ApJ 250, 155 }

\rf{Butchart, I., McFadzean, A. D., Whittet, D. C. B., Geballe, T. R., Greenberg, J. M. 1986, A\&A 154 }

\rf{Chan, Kin-Wing, Moseley, S. H., Casey, S., Harrington, J. P., 
    Dwek, E., Loewenstein, R., Varosi, F., Glaccum, W. 1997,
    ApJ 483, 798 }

\rf{Chiar, J.E., Tielens, A.G.G.M., Whittet, D.C.B., Schutte, W.A., Boogert, A.C.A., Lutz, D., van Dishoeck, E.F., Bernstein, M.P. 2000, ApJ 537, 749 }

\rf{Chiar, J. E., Tielens, A. G. G. M. 2001, ApJ 550, L207 }

\rf{Chiar, J. E., Adamson, A. J., Pendleton, Y. J., Whittet, D. C. B., 
    Caldwell, D. A., Gibb, E. L., 2002, ApJ 570, 198 }

\rf{Cl\'enet, Y., Rouan, D., Gendron, E., Montri, J.,
 Rigaut, F., L\'na, P., Lacombe, F. 2001, A\&A, 376, 124 }

\rf{Cl\'enet, Y. et al.,
   Proceedings of the Galactic Center Workshop, Nov. 3-8, 2002, Hawaii,
   A. Cotera, T. Geballe, S. Markoff, H. Falcke (editors), 2003,
   Astron. Nachrichten in press}

\rf{Cotera, A. S., Erickson, E. F., Colgan, S. W. J., Simpson, J. P., 
    Allen, D. A., \& Burton, M. G. 1996, ApJ 461, 750 }

\rf{Cotera, A. S., Erickson, E. F., Simpson, J. P., Rieke, M. 1992,
     J.American Astronomical Society, 180th AAS Meeting, 25.02, 
     Bulletin of the American Astronomical Society, Vol. 24, p.765 }

\rf{Cotera, A. S., Simpson, J. P., Erickson, E. F., Colgan, S. W. J., 
    Burton, M. G., Allen, D. A. 1999, ApJ 510, 747 }

\rf{Cotera, A., Morris, M., Ghez, A. M., Becklin, E. E., Tanner, A. M., Werner, M. W., Stolovy, S. R. 1999, cpg conf, 240 }

\rf{de Graauw, T., Whittet, D.C.B. et al. 1996, A\&A 315,L345}

\rf{Duley, W. W., Williams, D. A. 1983, MNRAS 205, 67 }

\rf{Duley, W. W., Williams, D. A. 1984, Natur 311, 685 }

\rf{Eckart, A. Moultaka, J. Viehmann, T. Straubmeier, C. Mouawad, N. 2004,
     ApJ 602, in press }

\rf{Eckart, A. Genzel, R., Hofmann, R., Sams, B.J. and Tacconi-Garman,
    L.E. 1995, ApJ 445, L26 }

\rf{Eckart, A. \& Genzel, R. 1996, Nature 383, 415-417 }

\rf{Eckart, A., Ott, T., Genzel, R., \& Lutz, D. 1998,
   in Proc. of IAU Symp. No.193 on 'Wolf-Rayet Phenomena
   in Massive Stars and Starburst Galaxies' Puetrto Valarta, Mexico,
   November 3--7, van der Hucht, K.A., Koenigsberger, G., 
   Enens, P.R.J.  (eds.), Kluewer, pp.449 }

\rf{Eckart, A, Ott, T, Genzel, R. 1999, A\&A 352,L22 }

\rf{Eckart, A., Genzel, R., Ott, T. and Schoedel, R. 2002, MNRAS 331,
   917-934 }

\rf{Eckart,A., Moultaka. J., et al., 2003, Proceedings of the Galactic Center Workshop, Nov. 3-8, 2002, Hawaii, A. Cotera, T. Geballe, S. Markoff, H. Falcke (editors, Astron. Nachrichten in press }

\rf{Eisenhauer, F., Schoedel, R., Genzel, R., Ott, T., Tecza, M., 
    Abuter, R., Eckart, A., Alexander, T., 
       accepret by ApJL (astro-ph/0306220)}

\rf{Figer, D.F., Gilmore, D., Morris, M., McLean, I.S., Becklin, E.E., Gilbert, A.M., Graham, J.R., Larkin, J.E., Levenson, N.A., Teplitz, H.I. 2001, AAS 198, 8706 }

\rf{Figer, D.F., et al. 2002, ApJ 581, 258 } 

\rf{Figer, D. F., Najarro, F., McLean, I. S., Morris, M., 
    Geballe, Th. R. 1997, Luminous Blue Variables: Massive Stars in Transition.
   ASP Conference Series, Vol. 120, 1997, ed. 
   Antonella Nota and Henny Lamers p.196 }

\rf{Figer, D.F., Najarro, F. et al. 2002, ApJ 581, 258 }

\rf{Fuente, A., Martin-Pintado, J., Bachiller, R.,
  Rodriguez-Franco, A., Palla, F. 2002, A\&A 387, 977 }

\rf{Genzel, R., Thatte, N., Krabbe, A., Kroker, H., Tacconi-Garman, L. E. 1996, ApJ 472, 153 }

\rf{Genzel, G., Eckart, A., Ott, T. \& Eisenhauer, F. 1997, MNRAS 291,
   219-234 }

\rf{Genzel, R., Pichon, C., Eckart, A., Gerhard, O. \& Ott, T. 2000,
 Mon.Not.R.Soc.317, 348-374 }

\rf{Genzel, R., Sch\"odel, R., Ott, T., et al. 2003, ApJ. 594, 812-832 }

\rf{Genzel, R., Sch\"odel, R., Ott, T., Eckart, A., Lacombe, F., Rouan,
  D., \& Aschenbach, B. 2003a, Nature 425, 934-936 }

\rf{Gerakines, P. A., Whittet, D. C. B., Ehrenfreund, P., Boogert, A. C. A., Tielens, A. G. G. M., Schutte, W. A., Chiar, J. E., van Dishoeck, E. F., Prusti, T., Helmich, F. P., de Graauw, Th. 1999, ApJ 522, 357 }

\rf{Gerakines, P. A., Whittet, D. C. B., Ehrenfreund, P., Boogert, A. C. A., Tielens, A. G. G. M., Schutte, W. A., Chiar, J. E., van Dishoeck, E. F., Prusti, T., Helmich, F. P., de Graauw, Th. 1999, ApJ 526.1062 }

\rf{Gerhard, O. 2001, ApJ 546, L39-L42 }

\rf{Gezari, D.Y., Shu, P., Lamb, G., Tresch-Fienberg, R., Fazio, G.G., 
    Hoffmann, W.F., Gatley, I., McCreight, C. 1985, ApJ 299, 1007 }

\rf{Gezari, D.Y., Schmitz, M., Pitts, P.S., Mead, J.M. 1993,
   Catalogue of Infrared Observations (NASA RP-1294)
   (3rd ed., Washington:NASA) }

\rf{Gezari, D., Dwek, E., Varosi, F.,
   In 'The Nuclei of Normal Galaxies: Lessons from the Galactic Center' 1994,
   Proceedings of the NATO Advanced Research Workshop, held in Schloss Ringberg, Kreuth, Bavaria, Germany, July 25-30, 1993, Dordrecht: Kluwer Academic Publishers, edited by Reinhard Genzel and Andrew I Harris. NATO Advanced Science Institutes (ASI) Series C, Volume 445, p.343 }

\rf{Gezari, D., Dwek, E., Varosi, F. 1996, IAUS 169, 231 }

\rf{Gezari, S., Ghez, A.M., Becklin, E.E., Larkin, J., McLean, I.S., Morris, M. 2002, ApJ 576, 790 }

\rf{Ghez, A., Klein, B.L., Morris, M. \& Becklin, E.E. 1998, ApJ. 509,
   678-686 }

\rf{Ghez, A., Morris, M., Becklin, E.E., Tanner, A. \& Kremenek, T. 2000, Nature 407, 349-351 }

\rf{Ghez, A., Duch\^ene, G., Matthews, K., et al. 2003, ApJ. 586,
   L127-L131 }

\rf{Gillett, F. C., Forrest, W. J. 1973, ApJ 179, 483 }

\rf{Glass, I.S., \& Moorwood, A.F.M. 1985, MNRAS 214, 429 }

\rf{Grishko, V. I., Duley, W. W. 2002, ApJ 568, L131 }

\rf{Guesten, R., Genzel, R., Wright, M. C. H., Jaffe, D. T., 
   Stutzki, J., Harris, A. I. 1987, ApJ 318, 124 }

\rf{Hagen, W., Greenberg, J. M., Tielens, A. G. G. M. 1983, A\&A 117, 132}

\rf{Herbst, T. M., Beckwith, S. V. W., Shure, M. 1993, ApJ 411, L21 }

\rf{Hillenbrand, L.A., Strom, S.E.,
  Vrba, F.J., Keene, J. 1992, ApJ 397, 613 }

\rf{Hoyle, F., Wickramasinghe, N. C., Al-Mufti, S., Olavesen, A. H., Wickramasinghe, D. T. 1982, Ap\&SS 83, 405 }

\rf{Hudgins, D. M., Sandford, S. A., Allamandola, L. J., Tielens, A. G. G. M. 1993, ApJS 86, 713}

\rf{Ishii, M., Nagata, T., Sato, S.,
   Watanabe, M., Y., Yongqiang, J., Terry J. 1998, AJ 116, 868 }

\rf{Jackson, J.M., et al. 1993, ApJ 402, 173 }

\rf{Jones, T. J., Hyland, A. R., Allen, D. A. 1983, MNRAS 205, 187 }

\rf{Joyce, R. R., Simon, T. 1982, ApJ 260, 604 }

\rf{Kim, S.S., Morris, M., Lee, H.M. 1999, ApJ 525, 228 }

\rf{Kitta, K., Kraetschmer, W. 1983, A\&A 122, 105}

\rf{Kleinmann, S.G., Hall, D.N.B. 1986, ApJS 62,501 }

\rf{Koornneef, J. 1983, A\&A 128, 84 }

\rf{Krabbe, A. et al. 1995, ApJL 447, L95 }

\rf{Lacy, J. H., Townes, C. H., Hollenbach, D. J. 1982, ApJ 262, 120 }

\rf{Lebofsky, M. J. 1979, AJ 84, 324 }

\rf{Lenzen, R., Hofmann, R., Bizenberger, P. \& Tusche 1998, A. Proc. SPIE, IR
    Astronomical Instrum.  (A.M.Fowler ed.) 3354, 606-614 }

\rf{Lutz, D. et al. 1996, A\&A 315, 269 }

\rf{Maillard et al.,
   Proceedings of the Galactic Center Workshop, Nov. 3-8, 2002, Hawaii,
   A. Cotera, T. Geballe, S. Markoff, H. Falcke (editors) 2003, Astron. Nachrichten in press }

\rf{Maldoni, Marco M., Smith, R. G., Robinson, Garry, Rookyard, V. L. 1998, MNRAS 298, 251}

\rf{Martin, P. G., Whittet, D. C. B. 1990, ApJ 357, 113 }

\rf{McFadzean, A.D., Whittet, D.C.B., Bode, M.F., Adamson, A.J., Longmore, A.J. 1989, MNRAS 241, 873 }

\rf{Mennella, V., Baratta, G. A., Esposito, A., Ferini, G., 
   Pendleton, Y. J. 2003, ApJ 587, 727}

\rf{Moneti, A., Cernicharo, J., Pardo, J.R. 2001a, ApJ 549, L203 .}

\rf{Morris, M., 1993, ApJ  408, 496.}

\rf{Najarro, F., Krabbe, A., Genzel, R., Lutz, D.,
  Kudritzki, R. P., Hillier, D. J. 1997, A\&A 325, 700 }

\rf{Ott, T., Eckart, A., Genzel, R., 2003, ApJ 523, 248O }

\rf{Paumard, T., Maillard, J.P., Morris, M., 
    Rigaut, F.  2001, A\&A 366, 466-480}

\rf{Pendleton, Y.J., Sandford, S.A., Allamandola, L.J., Tielens, A.G.G.M., Sellgren, K. 1994, ApJ 437, 683 }

\rf{Phinney, E.S., in The Centre of the Galaxy 1989, ed. M. 
    Morris (Dortrecht: Kluwer), 543 }

\rf{Portegies Zwart S., McMillan, S., Gerhard, O., 2003, astro-ph/0303599,
    ApJ in press.}

\rf{Reid, M.J., Menten, K.M., Genzel, R., Ott, T., Schödel, R., Eckart, A. 2003, ApJ 587, 208 }

\rf{Rieke, G.H., Low, F.G. 1973, ApJ 184, 415 }

\rf{Rieke, G. H., Lebofsky M.J. 1985, ApJ 288, 618 }

\rf{Rieke, G. H., Rieke, M. J., Paul, A. E. 1989, ApJ 336, 752 }

\rf{Rigaut et al.,
  Proceedings of the Galactic Center Workshop, Nov. 3-8, 2002, Hawaii,
   A. Cotera, T. Geballe, S. Markoff, H. Falcke (editors) 2003,
   Astron. Nachrichten in press }

\rf{Rousset, G. et al. 1998, Proc.SPIE Adaptive Optics Technology (D.Bonaccini \&
 R.K.Tyson eds) 3353, 508-516 }

\rf{Sanders, R.H. 1992, Nature 359, 131 }

\rf{Sandford, S.A., Allamandola, L.J., Tielens, A.G.G.M., Sellgren, K., Tapia, M., Pendleton, Y. 1991, ApJ 371, 607 }

\rf{Sandford, S. A., Pendleton, Y. J., Allamandola, L. J. 1995, ApJ 440, 697-705}

\rf{Scoville, N.Z., Stolovy, S.R., Rieke, M., Christopher, M.H., 
   Yusef-Zadeh F. 2003, accepted to ApJ (9/1/03 issue) }

\rf{Sch\"odel et al. 2002, Nature 419, 694-696 }

\rf{Sch\"odel, R., Ott, T., Genzel, R., Eckart, A., Mouawad, N., \&
  Alexander, T. 2003, Astrophys.J. 596, 1015-1034 }

\rf{Sellgren, K., McGinn, M. T., Becklin, E. E., Hall, D. N. 1990, ApJ 359, 112 }

\rf{Sellgren, K., Brooke, T. Y., Smith, R. G., Geballe, T. R. 1995, ApJL 449, L69 }

\rf{Serabyn, E., Morris, M. 1996, Nature 382, 602 }

\rf{Simon, M., Chen, W.J., Forrest, W.,J., Garnett J.D., Longmore, A.,J., Gauer, T., Dixon, R.I. 1990, ApJ 360, 95 }

\rf{Smith, R. G., Sellgren, K. \& Brooke, T. Y. 1993, MNRAS, 263, 749 }

\rf{Storey, J.W.V. \& Allen, D.A. 1983 MNRAS 204, 1153 }

\rf{Tamura, M., Werner, M.W., Becklin, E.E., Phinney, E.S. 1994, iaan conf. 117 }

\rf{Tamura, M., Werner, M. W., Becklin, E. E., Phinney, E. S. 1996, ApJ 467, 645 }

\rf{Tanner, A., Ghez, A. M., Morris, M., Becklin, E. E., Cotera, A., 
    Ressler, M., Werner, M., Wizinowich, P. 2002, ApJ 575, 860}

\rf{Tanner et al.,
   Proceedings of the Galactic Center Workshop, Nov. 3-8, 2002, Hawaii,
   A. Cotera, T. Geballe, S. Markoff, H. Falcke (editors),
   2003, Astron. Nachrichten in press }

\rf{Thatte, Niranjan A., Kroker, H., Weitzel, L., Tacconi-Garman, Lowell E., Tecza, M., Krabbe, Alfred, Genzel, R. 1995, SPIE. 2475, 228 }

\rf{Thornley, M.D., Schreiber, N.M. Förster, Lutz, D., Genzel, R., 
    Spoon, H.W. W., Kunze, D., Sternberg, A. 2000, ApJ 539, 641}

\rf{Tielens, A. G. G. M. \& Hagen, W. 1982, A\&A 114, 245}

\rf{Tielens, A. G. G. M., Hagen, W., Greenberg, J. M. 1983, JPhCh 87, 4220}

\rf{Tielens, A.G.G.M., Wooden, D.H., Allamandola, L.J., Bregman, J., Witteborn, F.C. 1996, ApJ 461, 210 }

\rf{Tokunaga, A. T., in 'Diffuse Infrared Radiation and the 
  IRTS' 1997, ASP Conference Series, Vol. 124, 1997, ed. H. Okuda, 
  T. Matsumoto, and T. Rollig, p.149 }

\rf{Tollestrup, E.V., Becklin, E.E., Capps, R. W. 1989, AJ 98, 204 }

\rf{Wada, S., Sakata, A., Tokunaga, A.T. 1991, ApJ 375, L17 }

\rf{Weitzel, L., Krabbe, A., Kroker, H., Thatte, N., Tacconi-Garman, L.E., Cameron, M., Genzel, R. 1996, A\&AS 119, 531 }

\rf{Willner et al. 1982, Ap.J. 253, 174 }

\rf{Yusef-Zadeh, F., Roberts, D. A., Biretta, J. 1998, ApJL 499, L159 }

\rf{Zhao, J.-H.,\& Goss, W.M. 1998, ApJL 499, L163 }

%\end{references}

\end{document}